\documentclass[aps,twocolumn,showpacs]{revtex4}
\usepackage{amssymb}
\usepackage{amsmath}
\usepackage{epsfig}
\begin{document}
\renewcommand{\theequation}{\thesection.\arabic{equation}}
\def\bq{\begin{equation}}
\def\eq{\end{equation}}
\def\bqa{\begin{eqnarray}}
\def\eqa{\end{eqnarray}}
\def\RR{\hbox{\it I\hskip -2.pt R }}
\def\bwt{\begin{widetext}}
\def\ewt{\end{widetext}}
\title{Quantum Fields as Operator Valued Distributions and Causality}
\author{ Pierre GRANG\'E}
\affiliation{Laboratoire de Physique Th\'eorique et Astroparticules\\
 Universit\'e Montpellier II, CNRS, IN2P3 \\
 CC 070 - Bat.13 - Place E. Bataillon\\
F-34095, Montpellier, Cedex 05 - FRANCE \\}
\author{Ernst WERNER}
\affiliation{Institut f\"ur Theoretische Physik\\
D-93040, Universit\"at Regensburg - GERMANY}
\date{\today}
\begin{abstract}
Quantum Field Theory with fields as Operator Valued Distributions 
with adequate test functions, -the basis of  Epstein-Glaser approach known now as 
Causal Perturbation Theory-, is recalled. Its recent revival is due to new developments in understanding 
its renormalization structure, which was a major and somehow fatal disease to its widespread
use in the seventies. In keeping with the usual way of definition of integrals of differential forms, 
fields are defined through integrals over the whole manifold, which are given an atlas-independent
meaning with the help of the  partition of unity. Using such partition of unity test functions
turns out to be the key to the fulfilment of the Poincar\'e commutator algebra as well as to provide
a direct Lorentz invariant scheme to the Epstein-Glaser extension procedure  of singular
distributions. These test functions also simplify the analysis of QFT behaviour both in the UV 
and IR domains, leaving only a finite renormalization at a point related to the
arbitrary scale present in the test functions. Some well known UV and IR cases are examplified. Finally 
the possible implementation of Epstein-Glaser approach in light-front field theory is discussed, focussing
on the intrinsic non-pertubative character of the initial light-cone interaction Hamiltonian and on the expected
benefits of a divergence-free procedure with only finite RG-analysis on physical observables in the end.
\end{abstract}


\maketitle

\setcounter{equation}{0}

\section{Introduction}
Causal perturbation theory (CPT) goes back to ideas from Stueckelberg, Rivier and Green \cite{Stu}, Bogoliubov, Parasiuk and 
Sirkov \cite{Bolo} and has been developed  by Epstein and Glaser \cite{EG73} and more recently by Scharf \cite{Scharf}. A little 
after Epstein and Glaser came formulations \cite{LZW} to construct finite gauge theories, based on regulator free momentum 
substraction schemes. 
 In the last publication the questions of the finiteness of Lorentz-invariant amplitudes and of symmetry conservation are treated
seperately. Moreover, as opposed to CPT, the importance of causality issues in relation to the finiteness problem itself is not
of main concern. In CPT the $S$-matrix is constructed as a formal functional expansion in an unimodular interaction-switching test function.
Whereas in the traditional approach the $S$-matrix amplitudes are determined by equations of motion, in CPT they are determined 
inductively by causality conditions imposed on switching test functions. Such causality conditions turn out to be as stringent for the 
$S$-matrix evolution as the equations of motion themselves. On an heuristic level it may be seen as follows: if  $g_{1}(t)$ and $g_{2}(t)$  
are two such switching test functions with supports such that $supp[g_{1}] \in (-\infty,s)$ and $ supp[g_{2}] \in (s,+\infty)$ respectively,
then causality implies that $S(g_{1}+g_{2})=S(g_{1})\cdot S(g_{2})$,  which is the functional equation for an exponential.\\
The $S$-matrix amplitudes are time ordered products of operator valued distributions (OPVD). They are split into causal advanced and 
retarded pieces, in such a way that any singular behaviour at equal space-time points is avoided. The procedure has long been recognized as 
mathematically rigorous \cite{Sto,ITZ} since it is divergence free at each iteration stage. Nevertheless Epstein-Glaser work almost
fell into oblivion among QFT practioners mainly because\\
(1) in its original version there were difficulties in disentangling the multiplicative structure of renormalization and \\
(2) the rise of renormalization group methods turned research interests into this direction.\\
\par
However Epstein-Glaser  approach still enjoys popularity in the world of mathematical physicists dedicated to the 
construction of a rigorous and mathematically well defined QFT \cite{BrFr,Mphys,GBL03}. In this context the key issue is the extension of 
singular distributions to the whole space-time manifold. In mathematics a rigorous way to define an extension of a singular distribution 
is a weighted Taylor series surgery: one throws away  an appropriate jet of the test function at the singularity and defines the extended 
distribution by  transposition at the level of the functional from the corresponding Taylor remainder of the test function. Transposed 
to Fourier space the procedure amounts to a substraction method \cite{GBL03} which includes
BPHZ renormalization \cite{Bolo,BPHZ} as a special case. In a Minkowskian metric this is equivalent to the implementation of causality 
while in its Euclidean counterpart it is a symmetry preserving prescription for substractions. In light-front quantum field theory (LFQFT) 
there is a compelling reason to introduce test functions related to the consistency of the canonical quantization scheme itself. It is best 
seen for the massive scalar field. The form of the  LF-Laplace operator leads to a hyperbolic equation of motion
which requires initial data on two characteristics. Canonical quantization in terms of initial field values in the light cone time is
 however possible provided
\cite{THEW} $\lim_{p^{+} \rightarrow 0}\frac{\chi(p^{+})}{p^{+}}=0$,where $\chi(p^{+})$ is the field amplitude at $p^{+}=p^{0}+p^{3}$. 
With fields as OPVD this relation becomes $\lim_{p^{+} \rightarrow 0 }\frac{f(p^{+})}{p^{+}}$, which is surely satisfied for the class of test functions 
$f(p^{+})$ used for the Fock expansion of the field operators \cite{GUW,ULL}.
 
Our earlier approach with test functions \cite{GUW,GSW} in fact implements Epstein-Glaser treatment of singular distributions
in the LFQFT context. Here we want to assess our particular choice  of  test functions as \lq partition of unity\lq\quad\!\!\!\!and show the resulting 
simplifications brought about for the treatment of Poincar\'e's invariance and for the Lorentz invariant Taylor series surgeries discussed
 in \cite{GBL03}.\\
In Section II the general definition of fields as OPVD is given. Due to specific properties of Euclidean manifolds it is shown that 
the class of admissible test functions can be restricted to a \lq partition of unity\lq, in close analogy with the usual atlas-independent 
definition of integrals of differential forms. Thus a QFT construction is possible, which is test-function independent. As a
consequence, for scalar field theory it is shown that Poincar\'e's commutator algebra is fulfilled, in contrast with the case of 
arbitrary test-functions.
In Section III the Bogoliubov-Epstein-Glaser expansion of the $S$-matrix is recalled. It is given in terms of the interaction-switching 
test function $g \in\mathcal{D}'(\RR^4)$. The proper Epstein-Glaser time ordered construction of the expansion coefficients is described.
The implementation of Lorentz covariance of the procedure is presented and the simplifications brought 
about with test functions vanishing at the origin with all their derivatives are shown. They concern Lorentz 
covariance, the Taylor surgery itself and its interpretation in terms of the BPHZ renormalisation scheme.
 In Section IV, the use of Lagrange's formula \cite{BrFr,GBL03} for Taylor's remainder in connection with partition of unity test functions
 is shown to simplify also the analysis of singular distributions both in the IR and UV domains. A direct application of the method in the
 IR is shown to be successful for the free massive scalar field theory developed in the (conventionally hopeless) 
mass perturbation expansion. It should also prove relevant to many problems of massless conformal field theories (CFT). Finally, following
\cite{PGEW06}, in the UV the alternative interpretation of the effects of partition of unity test functions in terms of 
Pauli-Villars type subtractions at the level of propagators is pointed out in Euclidean and Minskowskian metrics.
  In Section V Epstein-Glaser separation method of Section III for distributions into advanced and retarded pieces (splitting procedure) 
 is rephrased for the special case of partition of unity test functions. Their use simplifies greatly the derivation of the relation 
 between the splitting procedure and dispersion relations. In Section VI  a general discussion is given on the possibility of  
applying BSEG's method  in a non-perturbative Light Front Quantization framework and on the benefits one might expect from the procedure.
 Finally  some conclusions are drawn in Section VII.
\setcounter{equation}{0} 
\section{Fields as OPVD, partition of unity and Poincar\'e commutator algebra}
\subsection{Fields as OPVD}
The history of fields  as OPVD is almost as old as Quantum Field Theory itself. In fact the mathematical developments of Distribution Theory
(or Generalized Functions) \cite{LSCW,GUECHI} in the $1960's$ were intimately related to problems raised at that time  by pioneers in 
the QFT formulation. This mutual feedback has lead to QFT formulations \cite{BWH} in the $1960-70's$ which aimed at dealing 
in a mathematically consistent way with intrinsic divergences crippling the usual QFT perturbative analysis. Among these approaches the work
 of Epstein and Glaser \cite{EG73} has long been recognized as mathematically rigourous and free of undefined quantities. It is still 
 generating an important literature in mathematical physics and we shall summerise here some of its recent developments.\\
To introduce fields  as OPVD one may consider, without loss of generality, the free massive scalar field in D-dimensions. The general 
solution of the Klein-Gordon equation is a distribution, {\it{ie}} an OPVD, which defines a functional with respect to a test function $\rho(x)$
belonging to ${\mathcal D}(\RR^D)$, the space of Schwartz test functions \cite{LSCW},

\begin{equation}
\Phi(\rho) \equiv <\phi,\rho> = \int d^{(D)}y\phi(y)\rho(y).
\end{equation}
Here $\Phi(\rho)$ is an operator-valued functional with the possible interpretation of a more general functional 
$\Phi(x,\rho)$ evaluated at $x=0$. Indeed the translated functional is a well defined object \cite{LSCW} such that 
\begin{eqnarray}
\!\!\!\!T_{x}\Phi(\rho) &=& <T_{x}\phi,\rho> = <\phi,T_{-x}\rho> \\
\!\!\!\!&=& \int d^{(D)}y \phi(y) \rho(x-y).
\end{eqnarray}
Now the test function $\rho(x-y)$ has a well defined Fourier decomposition
\begin{equation}
\rho(x-y) = \int\frac{d^{(D)}q}{ (2 \pi)^{D}}\exp^{iq(x-y)}f(q_0^2,q^2).
\end{equation}
It follows that
\begin{equation*}
T_{x}\Phi(\rho) = \int \frac{d^{(D)}p}{ (2 \pi)^{D}}e^{-ipx}\delta(p^2-m^2)\chi(p)f(p_0^2,p^2).
\end{equation*}
Due to the properties of $\rho$, $ T_{x}\Phi(\rho)$ obeys the KG-equation and is taken as the physical field $\varphi(x)$. After 
integration over $p_0$ and with $\omega^2_{p}=p^2+m^2$ the quantized form is taken as the $(D-1)-$Euclidean integral  
\begin{equation}
\varphi(x) = \int\frac{d^{(D-1)}p}{ (2 \pi)^{(D-1)}}\frac{f(\omega^2_{p},p^2)}{2 \omega_{p}}[a^{+}_{p} e^{ipx} + a_{p} e^{-ipx}].
\end{equation}
$f(\omega^2_p,p^2)$ acts as a regulator \cite{GUW,ULL} with very specific properties \cite{foote1}. This expression for $\varphi(x)$ is particularly useful to define correlation functions of the field
and more precisely in the light-cone (LC) formalism because the Haag series can be used \cite{GSW}  and is well defined in terms of 
the product of $\varphi(x_i)$. In the full Euclidean metric there is no on-shell condition and $\varphi(x)$ stays a $D$-dimensional Fourier transform
 with $f(p^2)$ only.\\
 It might appear that there would be as many QFT's as eligible test functions.  However some importants results from topological spaces
 analysis may be invoked. Some general definitions need first to be recalled :\\
\par 
Definitions:\\
\par
a) An open covering of a topological space ${\mathcal M}$ is a family of open sets
$(U_i)_{i\in I}$ such that
${\mathcal M}=\cup_{i\in I }U_i$.\\
\par
b) An open covering  $(U_i)_{i\in I }$ is said to be locally finite if
each point of ${\mathcal M}$ has a neighbourhood which intersects only a finite
number of the $(U_i)$.\\
\par
c) A topological space ${\mathcal M}$ is paracompact iff every open covering
$(U_i)_{i\in I}$ admits a
locally finite refinement (that is a locally finite open covering
$(V_j)_{j\in J}$ with the property
that for all $j\in J$ there exists $i\in I$ such that $V_j\subseteq
U_i$).\\
\par
d) If  $(U_i)_{i\in I}$ is an open covering of ${\mathcal M}$, a  partition of
unity subordinate to $(U_i)_{i\in I}$ is a
family $(\beta_j)_{j\in J}$ of positive continuous functions on ${\mathcal M}$
such that\\

\hspace{0.5cm}(i) For all $X\in {\mathcal M}$ there exists a neighbourhood $U_X$ of $X$ such that
all but a finite number of the $\beta_j$ vanish on  $U_X$.\\

\hspace{0.5cm}(ii) $\sum_{i\in I} \beta_i(X)=1$ for all $X\in {\mathcal M}.$\\

\hspace{0.5cm}(iii) For all $j\in J$, there exists $i\in I$ such that the closure of
$\{X\in {\mathcal M}: \beta_j(X)\not=0\}$
is contained in $U_i$.  The closure of $\{X\in {\mathcal M}: \beta_j(X)\not=0\}$
is called the support of $\beta_j$.\\
\par
Remark: If  $(\alpha_i)_{i\in I}$  and $(\beta_j)_{j\in J}$ are  two
partitions of unity subordinate to
$(U_i)_{i\in I}$  and $(V_j)_{j\in J}$ respectively,
$(\alpha_i \beta_j)_{(i,j)\in I\times J}$ is a  partition of unity
subordinate to $(U_i\cap V_j)_{(i,j)\in I\times J}$. Since this property is crucial in the sequel it is established 
in Appendix A.\\
\par
Then the following theorems hold:\\
\par
Theorems:\\
\par
A)  Suppose ${\mathcal M}$ connected. Then ${\mathcal M}$ is paracompact iff the topology of
${\mathcal M}$ is metrisable iff (when ${\mathcal M}$ is a differentiable manifold) ${\mathcal M}$ admits a Riemannian
metric.\\
\par
B) If ${\mathcal M}$ is a connected,  paracompact differentiable manifold and
$(U_i)$ is an
open covering, then there exists a differentiable partition of unity
subordinate to $(U_i)$. \\
\par
Partitions of unity are useful because one can often use them to extend local
constructions to the whole space.  A well known case deals with the definition of integrals of differential forms 
\cite{DGEOM}: let $F$  be a differential form, $(\beta_i)$ is used to cut  $F$ into {\it small} pieces:  $F_i= \beta_i F $ 
and $\sum F_i=F$. For  $\Omega \subset {\mathcal M}$ one defines:  $\int_{\Omega} F := \sum_{i} \int_{\Omega_i} \beta_i F$. 
$\beta_i$ being of compact support each integral in the sum is finite and the overall result is independent of the choice 
of coordinates (atlas)  on $\Omega_i$ and  of the partition of unity.\\

C) Localisation of distributions: J. Dieudonn\'e's GPT-theorem (Glueing-Pieces-Together)\cite{LSCW,JDIEU}  establishes the above properties for
distribution functionals.\\ 
\par
The metrisable Euclidean manifold being paracompact permits the use of partition of unity test functions in
the definition of the field of Eq.(II.5). The resulting operator-valued functional is independent of the way this partition of unity 
is constructed. All constructions of the field with different partition of unity
 are therefore equivalent, thereby eliminating the initial arbitrariness in the choice of test functions.
 Then $f(p^2)$ is $1$ execpt in the boundary region where  it is infinitely differentiable and goes to zero  with all its 
 derivatives. An immediate and important consequence of this construction is that all successive convolutions of the field $\varphi(x)$ 
 with the test function leave $\varphi(x)$ unchanged. One has
\begin{equation}
<T_x \varphi,\rho> = \int \frac{d^{(D-1)}p}{(2 \pi)^{(D-1)}}\frac{f^2(\omega^2_{p},p^2)}{2 \omega_{p}}[a^{+}_{p} e^{ipx} + a_{p} e^{-ipx}].
\end{equation}
 Here $f^2$ stands for the product of two partitions of unity. According to the Remark above and Appendix A $f^2$ is
also a partition of unity with the same support as $f$  and $\varphi(x)$ is not affected.\\
\subsection{Partition of unity and Poincar\'e commutator algebra}
Without loss of generality, and to simplify the presentation, we consider here massive scalar fields at $D=4$ and use the following conventions:
$d\Omega_k\!=\!\frac{d^{3}k}{(2\pi)^3 2 \omega_k};\omega_k\!=\!\sqrt{k^2+m^2} ;[a_{\vec{k}},a^{+}_{\vec{k}'}]\!=\!(2\pi)^3 2 \omega_k
\delta(\vec{k}-\vec{k}')$. The energy-momentum tensor writes $\theta^{\mu\nu}=\partial^{\mu}\varphi \partial^{\nu}\varphi-\frac{1}{2}
g^{\mu\nu}[(\partial \varphi)^{2}-m^{2}\varphi^2]$ and gives
\begin{widetext}
\begin{eqnarray}
\!\!\!\!\!\!\!\!\!\!& & P^0=H=\int d\Omega_k \omega_k a^{+}_{\vec{k}}a_{\vec{k}}f^{2}(\omega^{2}_{k},\vec{k}^2);\quad \!\!\!\!
P^j=\frac{1}{2}\int d\Omega_k \omega_k k^{j}[a^{+}_{\vec{k}}a_{\vec{k}}+a_{\vec{k}}a^{+}_{\vec{k}}]f^{2}(\omega^{2}_{k},\vec{k}^2);\\
\!\!\!\!\!\!\!\!\!\!& & M_{0j}=i\int d\Omega_k \omega_k a^{+}_{\vec{k}} f(\omega^{2}_{k},\vec{k}^2)
\frac{\partial}{\partial k^j}(a_{\vec{k}} f(\omega^{2}_{k},\vec{k}^2));\quad \!\!\!\! 
M_{jl}=i\int d\Omega_k  a^{+}_{\vec{k}} f(\omega^{2}_{k},\vec{k}^2)
[k_j\frac{\partial}{\partial k^l}-k_l\frac{\partial}{\partial k^j}](a_{\vec{k}} f(\omega^{2}_{k},\vec{k}^2)).
\end{eqnarray}
\end{widetext}
Evaluating the commutator of $P^j$  with $\varphi(x)$ gives in turn
\begin{equation}
i[P^j,\varphi(x)] =i\int d\Omega_k 
k^{j}[a^{+}_{\vec{k}}e^{ikx}-a_{\vec{k}}e^{-ikx}]f^3(\omega_k^{2},\vec{k}^2),
\end{equation}
which is the expected result $\partial^j \varphi(x)$ because $f^3$ is also an equivalent partition of unity. For all other commutators
 the reduction procedure due to the $a$'s and $a^+$'s is the usual one and since any power of the test function is an equivalent
partition of unity, the Poincar\'e algebra is thereby satisfied. It is clear that any arbitrary test
 function will fail in that respect.
\setcounter{equation}{0}
\section{Bogoliubov-Shirkov-Epstein-Glaser (BSEG) expansion of the $S$-matrix and Lorentz invariant extension of singular distributions}
\subsection{The BSEG procedure}
In the formal expansion of the $S-$matrix \cite{Bolo} the coefficients are OPVD built out of products of free fields. According to the above
construction products of test functions appear then naturally and in the BSEG writing of $S$ they are made explicit
\begin{equation}
S(g)\!=\!1\!+\!{\displaystyle\!\!\sum_{n=1}^{\infty}}\frac{1}{n!}\int\!\!d^4x_1..d^4x_nT_n(x_1,..,x_n)g(x_1)..g(x_n).
\end{equation}
The test functions $g(x)$ must have two properties:\\
\par
-$i)$ vanish at infinite time in order to switch off the interaction. This is necessary to define asymptotically free
states. It is a problem in equal time QFT because non-trivial vacua are excluded, but in LCQFT, because of the LC-vacuum structure, one can 
build a non perturbative theory on free field operators.\\
\par
-$ii)$ restrict the propagation of fields to the interior of the light cone.\\
\par
Suppose that all time arguments of the set of vectors $\{x_1,\cdots,x_m\}$ are bigger than those of the set $\{x_{m+1},\cdots,x_n\}$ then the
supports of the ensemble of test functions $\{g(x_1),\cdots,g(x_m)\}$ are all separated from those of $\{g(x_{m+1},\cdots,g(x_n)\}$ . The 
requirement of causality leads to
\begin{equation}
T_n(x_1,\cdots,x_n)=T_m(x_1,\cdots,x_m)T_{n-m}(x_{m+1},\cdots,x_n).
\end{equation}
As already mentioned in the introduction this condition completely fixes the dynamics of the system. If all arguments are different and
ordered such that $x_1^0>x_2^0>\cdots>x_n^0$, by recursion it follows that
$T_n(x_1,\cdots,x_n)=T_1(x_1)T_1(x_2)\cdots T_1(x_n)$ and 
\begin{equation}
S(g)\!=\!1+\!{\displaystyle \sum_{n=1}^{\infty}}\frac{1}{n!}\!\!\int\!\!d^4x_1..d^4x_n\mathcal{T}[T_1(x_1)..T_1(x_n)]g(x_1)..g(x_n).
\end{equation} 
It is important to note that the time ordering operation $\mathcal{T}$ cannot be made with $\theta$-functions because the multiplication of
the two distributions $\theta(x)$ and $T_i(x)$ at the same point is mathematically ill-defined. Epstein-Glaser analysis focusses then 
on the solution of this difficulty to which we now turn.
\subsection{Epstein-Glaser extension of singular distributions} 
Because the functional integration of Eq.(II.1) may occur in practice in a space of smaller dimension than the original dimension $D$
we shall use $d$ as a dimensional label henceforth. Due to translation invariance the multiplication problem just mentionned can be reduced 
to the study of distributions singular at the origin of $\RR^{d}$. Let then  
$f(X)$ be a ${\mathbb C}^{\infty}$ test function belonging to ${\mathcal D}(\RR^{d})$ 
and $T(X)$ a distribution belonging to ${\mathcal D}^{\prime}(\RR^{d}\backslash\{0\})$ which we want to extend to the whole domain ${\mathcal
D}^{\prime}(\RR^{d})$. The singular order  $k$  of $T(X)$   at the origin of $\RR^{d}$  is defined as 
\begin{equation}
k=inf \{s:{\displaystyle \lim_{\lambda \rightarrow 0}}\lambda ^{s} T(\lambda X) = 0\} - d.
\end{equation}
 Epstein-Glaser original extension consists in performing an \lq educated\lq\quad\!\!\!Taylor surgery on the initial test function by throwing away 
 the  weigthed k-jet of $f(X)$ at the origin. Denoting by $R^{k}_{1}f(X)$ this Taylor's remainder, this implies 
\begin{equation}
 R^{k}_{1}f(X)=f(X)-\sum_{n=0}^{k}\sum_{|\alpha|=n}\frac{ X^{\alpha}}{\alpha!}\partial^{\alpha}f(X)|_{X=0}.                                       
\end{equation} 
 Here the notations are : $\alpha!=\alpha^1!\cdots\alpha^d!$, $|\alpha|=\alpha^1+\cdots+\alpha^d$ and $\partial^\alpha=\partial^{\alpha^1}_{x_1}\cdots\
 \partial^{\alpha^d}_{x_d}$. $R^{k}_{1}f(X)$ has the desired properties at the origin  -it vanishes there to order $k+1$-  but, due to
 uncontrolled UV-behaviour, it does not belong to Schwartz-space ${\mathcal D}(\RR^{d})$. In order to be able to define a functional 
 which can be used to perform a transposition in the sense of distributions Epstein and Glaser introduce a weight function  $w(X)$
 belonging to ${\mathcal D}(\RR^{d})$ with properties \quad $w(0)=1$,\quad $w^{(\alpha)}(0)=0$ \quad for \\ $0 <\mid \alpha \mid\leq k$. This allows to define a
 modified Taylor remainder {\it {eg}}:
\begin{equation}
R^{k}_{w}f(X)=f(X)-w(X)\sum_{n=0}^{k}\sum_{|\alpha|=n}\frac{ X^{\alpha}}{\alpha!}\partial^{\alpha}f(X)|_{X=0}.
\end{equation}
Now the extension $\widetilde{T}_{w}(X)$ of $T(X)$ can be defined by transposition:
\begin{equation}
<\widetilde{T}_{w},f>:=<T,R^{k}_{w}f>.
\end{equation}
The IR-extension  $\widetilde{T}_{w}(X)$ so obtained is not unique. We shall come back to this point . There is an abundant
literature on this original procedure of Epstein-Glaser and some important improvements were proposed recently \cite{Mphys,GBL03}.
Our observation here \cite{PGEW06} concerns the use of test functions belonging to ${\mathcal D}(\RR^{d})$  and vanishing at the origin 
of $\RR^{d}$ with all their derivatives (dubbed super-regular or SRTF hereafter). In this case one has strictly $f(X)=R^{k}_{1}f(X)$, for all the 
derivative terms in Eq.(III.5) are zero. Then $R^{k}_{1}f(X)$ belongs also to ${\mathcal D}(\RR^{d})$, there is no need to introduce a weight 
function and the following set of equalities holds
\begin{equation}
 <T,f>=<T,R^{k}_{1}f>=<R^{k}_{1}T,f>\equiv<\widetilde{T},f>.
\end{equation}
 According to Section II $f$  stands here for a partition of unity which allows defining $<T,f>$ from theorem C. Then,  
 $\widetilde{T}$ becomes a regular distribution over the whole manifold on which  $f$ is just $1$ everywhere. The way this is achieved 
 in practice will be shown in Section IV.
Important consequences follow from these relations. They have to do with the splitting procedure of
distributions into advanced and retarded parts, Lorentz covariance of the Epstein-Glaser analysis, the connection with BPHZ renormalization 
and consequently the analysis of the IR and UV behaviour of the underlerlying QFT.\\
\subsection{Splitting into advanced and retarded parts of singular distributions}
As mentionned earlier the time ordering operation $\mathcal{T}$ in Eq.(III.3)  cannot be made blindly without facing divergences related to 
 ill-defined products of distributions. This ordering operation requires to split distributions $T$ into advanced and retarded parts $T_a$ 
 and $T_r$ respectively in the following way
\begin{equation}
 T(x_1,\cdots,x_n)=T_r(x_1,\cdots,x_n)-T_a(x_1,\cdots,x_n).
\end{equation}
The support of $T_r(x_1,\cdots,x_n)$ ({\it {viz.}}\quad\!\!\!\! $T_a(x_1,\cdots,x_n)$ )is the n-dimensional generalization of the closed foreward (backward)
cone of $\{x_1,\cdots,x_n\}$. Because of translational invariance one can put $x_n=0$. If $T(x_1,\cdots,x_n)$ were regular at
$x_i=0,i=1,\cdots,n-1$ the splitting could be performed with
\begin{eqnarray}
T_r(x_1,..,x_n)\!\!&=&\!\!\!\prod_{i=1}^{n-1}\!\!\theta(x_j^0-x_n^0)T(x_1-x_n,..,x_{n-1}-x_n,0),\nonumber \\
T_a(x_1,..,x_n)\!\!&=&\!\!\!\prod_{i=1}^{n-1}\!\!\theta(x_n^0-x_j^0)T(x_1-x_n,..,x_{n-1}-x_n,0).\nonumber \\
\end{eqnarray}
On the contrary, if the product of $\theta$-functions with the distribution $T$ is ill-defined the splitting procedure has to be done with
the help of Eq.(III.8). The defining equations of $\widetilde{T}_r$ and $\widetilde{T}_a$ are then
\begin{eqnarray}
\!\!\!\!<T,\theta f>\!\!\!&=&\!\!\!<\theta \widetilde{T},f>=<\widetilde{T}_r,f>; \\
\!\!\!\!<T,(1-\theta) f>\!\!\!&=&\!\!\!<(1-\theta )\widetilde{T},f>=<\widetilde{T}_a,f>.\nonumber
\end{eqnarray}
$\widetilde{T}_r$ and $\widetilde{T}_a$ are simply obtained by multiplication of $\widetilde{T}$ with the corresponding
$\theta$-functions. With the original singular distribution $T$ this would not have been possible. Finally the following identification is
obtained
\begin{equation}
\widetilde{T}_r=\theta \widetilde{T} ;\quad  \widetilde{T}_a=(1-\theta) \widetilde{T}.
\end{equation}
It should be kept in mind that  $\widetilde{T}_r$ and $\widetilde{T}_a$ are not unique, since, as stated above, $\widetilde{T}$ is not
unique.\\
If $T(X)=T(x_1,\cdots,x_n)$ is causal -as is the case, if it is constructed with the BSEG procedure sketched in Section (III.A)- then the
products of $\theta$-functions in Eq.(III.10) allow an essential simplification particularly useful for calculations in momentum space
\cite{Scharf}. There Scharf defines a $4(n-1)$-dimensional vector $v=(v_1, v_2, \cdots, v_{n-1})$ consisting of $n-1$ time-like four-vectors
lying all in the interior of the foreward light-cone. The scalar product $v_i\cdot X_i=v_i^0 X_i^0-\overrightarrow{v} \cdot \overrightarrow{X}$
is used to define the scalar  $v\cdot X=\sum_{j=1}^{n-1} v_j \cdot X_j$, where one puts $X_n=0$. Then $v\cdot X$ has the following properties:\\
\par
$v\cdot X > 0$,\quad $\forall X_j$ in the foreward light-cone $\Gamma^+$\\
\par
$v\cdot X < 0$,\quad $\forall X_j$ in the backward light-cone $\Gamma^-$\\
\par
Consequently $v\cdot X =0$ defines an hyperplane that separates the causal support into advanced and retarded parts.\\
The following indentities then hold:
\begin{eqnarray}
\theta(v\cdot X)&=&\prod_{j=1}^{n-1}\theta(X_j^0-X_n^0);\quad X \in \Gamma^+ \nonumber \\
\theta(v\cdot X)&=&\prod_{j=1}^{n-1}\theta(X_n^0-X_j^0);\quad X \in \Gamma^- 
\end{eqnarray}
These identities can only be used with distributions having causal supports. Otherwise the scalar products $v_i\cdot X_i$ would not have a
unique sign related to advanced or retarded coordinates and $v\cdot X=0$ could not be used to distinguish advanced and retarded parts of the
supports. For later use we need the Fourier transform of $\theta(v\cdot X)$, {\it i.e.} $\overline{\theta_v}(p)$ where $p=(p_1, p_2, \cdots,
p_{n-1})$. This multi-dimensional quantity can be reduced to a $4$-dimensional one by\\
\par
$(1)$ choosing a coordinate system for which $p=(p_1,0,\cdots,0)$ where $p_1=(p_1^0,\overrightarrow{p_1})$ (which can be obtained
by an orthogonal transformation in $\RR^{4(n-1)}$).\\
\par
$(2)$ and choosing $v=(v_1,0,\cdots,0)$; $v_1=(1,\overrightarrow{v})$  which yields $\theta(v\cdot X)=\theta(X_1^0-
\overrightarrow{X_1}\cdot\overrightarrow{v})$. Then\\
\par
\begin{eqnarray}
\!\!\!\!\!\!\!\!\overline{\theta_v}(p)=\frac{1}{(2\pi)^{D/2}}\int\!\!d^{D}X_1 e^{ip.X_1}\theta(X_1^0-\overrightarrow{X_1}\cdot\overrightarrow{v})
\end{eqnarray}
Now with
\begin{equation}
\theta(X_1^0-\overrightarrow{X_1}\cdot\overrightarrow{v})=\frac{1}{2\pi i}\int_{-\infty}^{\infty}d\tau \frac{e^{i\tau(X_1^0-
\overrightarrow{X_1}\cdot\overrightarrow{v})}}{\tau-i\epsilon},\nonumber
\end{equation}
one obtains:
\begin{equation}
\overline{\theta_v}(p)=(2\pi)^{(D/2-1)}\frac{1}{p^0+i\epsilon}\delta^{(D-1)}(\overrightarrow{p}-p^0\overrightarrow{v}).
\end{equation}
The last equality  means that $\overrightarrow{p}$ and $\overrightarrow{v}$ come out parallel to each other. Of course physical results
should not depend upon the choice of the arbitrary vector $v$. This will be verified in Section V.
\subsection{Lorentz invariance of Epstein-Glaser procedure and super-regular test functions. } 
In the Taylor surgery above, under the action of an element $\Lambda$ of the Lorentz group, derivatives transform as
\begin{eqnarray}
 x^{\alpha}\partial_{\alpha}(\Lambda f)&=& x^{\alpha}[\Lambda^{-1}]^{\beta}_{\alpha}(\partial_{\beta}f) \circ \Lambda^{-1}, \nonumber \\
 &=&(\Lambda^{-1} X)^{\beta}(\partial_{\beta} f) \circ  \Lambda^{-1},
\end{eqnarray}  
  and in the general Taylor expansion $x^{\alpha}\partial_{\alpha}(\Lambda f)(0)=(\Lambda^{-1} X)^{\beta}\partial_{\beta}f(0)$.
Lorentz invariance is therefore violated in this procedure. However, as mentionned earlier, $\widetilde{T}(X)$ is determined up to a sum of
derivatives of $\delta$ functions  $\displaystyle \sum_{\mid \alpha \mid\leq k}
(-1)^{\alpha}\frac{a_{\alpha}}{\alpha!}\delta^{(\alpha)}(X) $. But these $\delta$-terms transform as $\delta^{(\alpha)}(\Lambda X)=[\Lambda^{-1}
X]^{\alpha}_{\beta}\delta^{(\beta)}(X)$. Epstein-Glaser remedy is then to determine the  $a_{\alpha}$'s to correct for the violation due
 to derivatives. With SRTF vanishing at the origin with all their derivatives, because of the identity 
$f(X)\equiv R^{k}_{1}f(X)$, Lorentz violating derivatives are just not there. Provided then that the space of test functions to be used is 
restricted to SRTF types, Lorentz invariance is satisfied from the outset. Formally  $\widetilde{T}(X)$ is only determined
 up to the sum of derivatives of $\delta$-terms  but their contributions are redundant with super-regular test functions. Nevertheless its presence should not be
 forgotten and may prove instrumental in restoring some other broken symmetries \cite{Scharf}.
\subsection{ BPHZ renormalization: a corollary of Epstein-Glaser procedure.}
One of the obstacles for a widespread application of the Epstein-Glaser approach was the non evident link to other renormalization
schemes and in particular the question of its multipicative structure and its feasability in renormalization group studies. The chain of
identities of Eq.(III.8) valid for SRTF allows to  establish quite simply the link of our approach with BPHZ renormalization. In the 
context of the Ces\`aro interpretation of diverging integrals \cite{Kanw} this link has been discussed by Prange \cite{Mphys} and
Gracia-Bondia \cite{GBL03}. In the approach with SRTF the link to BPHZ is obtained as follows.\\
Fourier and inverse Fourier transforms are defined as
\begin{eqnarray}
\!\!\!\!\!\!\!\!{\bf {\mathbb F}}(p)&=&{\mathcal F}[f](p)=\int\frac{d^{d}X}{(2 \pi)^{d / 2}}e^{-ipX}f(X); \nonumber \\ 
\!\!\!\!\!\!\!\!\overline {{\bf {\mathbb F}}}(p)&=&{\mathcal F}^{-1}[f](p)=\int\frac{d^{d}X}{(2 \pi)^{d / 2}}e^{ipX}f(X). 
\end{eqnarray}
Then, with $\mu$ a multi-index as defined after Eq.(III.5), the following relations
hold
\begin{eqnarray}
&&\overline {( X^\mu f)}(p)=(-i)^{|\mu|} \partial^\mu \overline {{\bf {\mathbb F}}}(p), \nonumber \\
&&\partial^\mu f(0)=(-i)^{|\mu|} (2 \pi)^{-d/2}<p^\mu,\overline {{\bf {\mathbb F}}}>, \nonumber \\
&&\overline {( X^\mu)}(p)=(-i)^{|\mu|} (2 \pi)^{d/2}\delta^{(\mu)}(p).
\end{eqnarray}
The Fourier transform of the singular $T(X)$ is well defined since the functional built with SRTF
identical to their Taylor remainder is itself well defined. Therefore relations (III.8) and (III.18) imply
\begin{eqnarray}
<T,f>&\equiv&<T,R^{k}_{1}f>=<{\mathcal F}(T),{\mathcal F}^{-1}(R^{k}_{1}f)> \nonumber\\
&=&<{\mathcal F}(T),R^{k}_{1}({\mathcal F}^{-1}(f))>\nonumber \\
&=&<R^{k}_{1}{\mathcal F}(T),{\mathcal F}^{-1}(f)>,
\end{eqnarray}
that is
\begin{equation}
{\mathcal F}({\tilde T})=R^{k}_{1}{\mathcal F}(T).
\end{equation}
This is BPHZ substraction at zero momentum, up, as is well known, to an arbitrary polynomial in p originating from the sum of $\delta$-terms
mentioned previously. It is known that for a non-massive QFT the BPHZ method faces infrared divergences. However, a mass scale can be introduced
 by doing substractions at some external momenta $q\not= 0$. For non-SRTF it amounts to choosing the weight function 
 $w(X)=e^{iqX}$. This is a legitimate choice provided the functional integral in Eq.(III.7) is given a meaning in the sense of Ces\`{a}ro summability 
\cite{GBL03,Kanw}.
 However the situation is much simpler with SRTF. One has still 
$e^{iqX}f(X)\equiv R^{k}_{1}(e^{iqX}f(X))$ and the chain of equalities in Eq.(III.19) can be rewritten with this modification of $f$, leading now 
to BPHZ substraction at momentum $p=q$. Thus BPHZ appears just as a special case of Epstein-Glaser method. In the next section we examine
 in more details UV and IR behaviours when using partition of unity test functions introduced in Section II.
\setcounter{equation}{0}
\section{IR and UV extensions of $T(X)$ with Lagrange's formula for Taylor's remainder and partition of unity} 
\subsection{Lagrange's formula in the IR}
It is known from general functional analysis that Taylor's remainder can be expressed by Lagrange's formula. It is only recently \cite{BrFr}
that its use was advocated in the Epstein-Glaser context of extension of singular distributions and in the ensuing
 RG-analysis. With Lagrange's formula the transposition operation defining the extension $\widetilde{T}(X)$ reduces 
 to partial integrations in the functional integral. According to Lagrange $R^{k}_{1}f(X)$ can be written as
\begin{equation}
R^{k}_{1}f(X) =(k+1){\displaystyle \sum_{\mid \beta \mid=k+1}}\big[
\frac{X^{\beta}}{\beta !}\int^{1}_{0} dt (1-t)^{k}\partial^{\beta}_{(tX)}f(tX)\big].
\end{equation}
 For a SRTF $\in {\mathcal D}(\RR^{d})$ it is easily verified that for any $k$ Eq.(IV.1) expresses the identity
 $f(X)\equiv R^{k}_{1}f(X)$. Since this identity is also valid for any power of $f(X)$,  the field construction and  
 Poincar\'e commutator algebra keep their initial properties with respect to the partition of unity (section II-(A,B)).
One has then
\bwt
\begin{equation}
<T,f>=<T,R^{k}_{1}f>=(k+1){\displaystyle \sum_{\mid \beta \mid=k+1}}\int d^{d}X T(X)\big[
\frac{X^{\beta}}{\beta !}\int^{1}_{0} dt (1-t)^{k}\partial^{\beta}_{(tX)}f(tX)\big].
\end{equation}
\ewt
Performing partial integrations in $X$, surface terms are avoided with SRTF. This is necessary  to
 validate the usual operations with distributions. It turns out from the generic features of $f(X)$ \cite{PGEW06}, 
detailed in Appendix B, that the t-integral has an effective lower cut-off at $\tilde{\mu}\| X\|<1$  for $\tilde{\mu}<1$, where 
$\| X\|<1$ is the norm of $X$. 
The extension is looked for in the IR region of the variable $X$. Since in the following $X$ can be a spatial or a momentum
 variable "infrared" in the present context can mean UV (small distance) or IR (large distance) in physical terminology. Since for a 
 momemtum variable the extension $\widetilde{T}^{<}(X)$ is infrared and/or ultraviolet regular by construction, the test function can now
  be replaced by $1$ over the whole domain of integration \cite{PGEW06}. One obtains:
\bwt
 \begin{eqnarray} 
\widetilde{T}^{<}(X)&=&(-)^{k+1}(k+1){\displaystyle \sum_{\mid \beta \mid=k+1}}
 \partial^{\beta}_{X}\big[\frac{X^{\beta}}{\beta !}\int_{\tilde{\mu}\| X \|}^{1}dt \frac{(1-t)^{k}}{t^{(k+d+1)}}T(\frac{X}{t})\big].
\end{eqnarray}
\ewt
For an homogeneous distribution, that is $T(\frac{X}{t})=t^{(k+d)}T(X)$, the $t$-integration can be carried out to give 
\bwt
\begin{eqnarray}
\!\!\!\!\!\!\widetilde{T}^{<}(X)&=&(-)^{k}(k+1){\displaystyle \sum_{\mid \beta \mid=k+1}} \partial^{\beta}_{X}
\big[\frac{X^{\beta}}{\beta !} T(X)\log(\tilde{\mu}\| X \|)\big]+\frac{(-)^{k}}{k !}H_{k}{\displaystyle \sum_{\mid \beta \mid=k}}
 C^{\beta} \delta^{(\beta)}(X), 
\end{eqnarray}
\ewt
with $H_{k}={\displaystyle\sum_{p=1}^{k}}\frac{(-1)^{(p+1)}}{p}\left(\begin{array}{c}k\\p\end{array}\right)=\gamma+\psi(k+1)$ and
$C^{\beta}=\int_{(\| X \|=1)} T(X) X^{\beta} dS$.\\
It is interesting to note how this result \cite{foote2} parallels that of Ref.\cite{GBL03}. Here it is the behaviour of the SRTF
at the origin which provides the lower bound in the t-integral and at the same time gives the identity $f(X)\equiv
(1-w(X))R^{k-1}_{1}f(X)+w(X)R^{k}_{1}f(X)\quad \forall w(X)$. It just corresponds to the $T_{w}$ operation  of \cite{GBL03} applied 
on any test function $\phi(X) \in {\mathcal D}(\RR^d)$ with the choice $w(X)=\theta(1-\tilde{\mu}\|X\|)$.\\
\par
The power of the IR treatment through Eq.(IV.4) is shown in \cite{GBL03,PGEW06} for the free massive scalar field theory developed in the
 (conventionally hopeless) mass perturbation expansion. It leads through the exact resummation of the infinite mass-expansion series to the well
 known results given in terms of modified Bessel functions $K_\nu(m \| X \|)$. Another interesting test concerns  integrable conformal field
 theories with additive interacting terms of massive dimension, which endure also conventionally untractable IR divergences if treated perturbatively
 \cite{LuZa}.  
\subsection{Extension of $T(X)$ in the UV domain}
From the above IR analysis one could also explore the Fourier space UV domain. However the same formalism can be directly applied if $T(X)$
gives rise to UV divergences in the absence of test functions. In QFT $T(X)$ is in general a distribution in the variable $\|X\|$ only.
Setting $\|X\| \equiv  X$ from now on, the domain of $f(X)$ is the ball  $B_{1+h}(\| X \|)$  of radius $1+h$ and $f^{(\alpha)}(1+h)=0,\forall \alpha \geq 0$. With
 $f(X) \equiv f^{>}(X)$, it also holds that
\begin{eqnarray}
\!\!\!\!\!\!f^{>}(X)\!\!&\equiv&\!\!-\frac{1}{k!}\int^{\infty}_{1} \!\!\!\!dt (1-t)^{k}\partial^{(k+1)}_{t}\big[t^k f^{>}(Xt)\big],\nonumber \\
\!\!\!\!\!\!&=&\!\!-\frac{X}{k!}\!\!\int^{\infty}_{1} \!\!\!\frac{dt}{t}(1-t)^{k}\partial^{(k+1)}_{X}\big[X^k f^{>}(Xt)\big].
\end{eqnarray}
Whatever the construction of the partition of unity underlying $f(X)\quad h$ is a parameter which may depend on $\| X \|$. The arguments
underlying this observation are as follows. With $h$ taken as a true constant the regularisation will be achieved in the usual 
cut-off like  manner with well known symmetry 
preserving problems. In this case the final extension of the partition of unity to the whole manifold
can only take place by finally letting $h$ -the cut-off- going to infinity
thereby losing the scaling information originally contained in
the construction of the overlapping domains. Clearly using a partition of
unity would be a useless artifact in this case. However with an
$\| X \|$-dependant $h$, as we shall see, the situation is very different:
the final extension of the partition of unity to the whole
manifold can be achieved altogether in a manner preserving symmetries and
scaling informations. With an $\| X \|$-dependant $h$ the consequences are:\\
\par
$-i)$ $\exists$    $\|X\|_{max}$   such that   $\|X\|_{max}=1+h(\|X\|_{max})\equiv 
\mu^2 \|X\|_{max} g(\|X\|_{max})$ $\Longrightarrow$ $g(\|X\|_{max})=\frac{1}{\mu^2}$,\\
\par
$-ii)$   $h >0$ $\Longrightarrow$ $\mu^2\|X\| g(\|X\|) > 1 $ $\forall$  $\|X\|$ $\in$  $[1,\|X\|_{max}]$ 
$\Longrightarrow$  $g(1) > g(\|X\|_{max})$, \\
\par 
$-iii)$ from $f^{>}(tX)$ present in Lagrange's formula
one has  $t < \frac{1+h(\|X\|)}{\|X\|}= \mu^2 g(\|X\|)$. \\
\par
In the definition of $h(\|X\|)$ a dimensionless scale factor $\mu^2$ has been extracted from $g(\|X\|)$. Then one has
\bwt
\begin{eqnarray}
\!\!<T,f^{>}>\!\!&=&\!\!\int \!\!d^{d}X  T(X)\big{\{}-\frac{X}{k!}\int^{\!\!\ ^{^{\mu^2 g(\|X\|)}}}_{1} \!\!\!\!\!\!\!\!\frac{dt}{t}(1-t)^{k}\partial^{(k+1)}_{X}\big[X^k
f^{>}(Xt)\big]\big{\}} \nonumber \\
 &=&<\widetilde{T}^{>},1>, 
\end{eqnarray}
\ewt
where in the last line the partial integrations in $X$ have been performed. It gives the $UV$-extension $\widetilde{T}^{>}(X)$ of $T(X)$ over
the whole $X$ domain on which $f$ is just one everywhere. .
\begin{equation}
\widetilde{T}^{>}(X)\!=\!\frac{(-)^{k}}{k!} X^{(k-d+1)}\partial^{(k+1)}_{X} \big[X^d T(X)\!\!\int^{\!\!\ ^{^{\mu^2 g(\|X\|)}}}_{1} \!\!\!\!\!\!\!\!   
\frac{dt}{t}(1-t)^{k}\big].
\end{equation}
A direct application of this relation in the Euclidean metric is made for the scalar propagator $\Delta(x-y)$ which is diverging
 for $D=2\cdots 4$ when $x=y$. These cases are important to clarify the role of the presently unknown function $g(\|X\|)$, on the one hand in
producing a regular UV-extension $\widetilde{T}^{>}(X)$ and, on the other hand, in the final RG-analysis with respect to the scale parameter 
$\mu$ present in $f(X)$. Suppose now that $T(X)\sim \|X\|^{-(\omega+1)}$ as $\|X\| \rightarrow \infty$, then $\int d^{(d)}X  T(X)$ diverges if
$(\omega +1-d) \leq 0$. Thus $k$ in Eq.(IV.7) should be such that $k \geq (d-\omega-1)$.
Consider the Euclidean scalar propagator at $x=y$ and for $D=4$.  The argument of the test function is dimensionless. Hence we
 have $X=\frac{p^2}{\Lambda^2}$, where $\Lambda$ is an arbitrary scale which will prove irrelevant. Thus $T(X)=\frac{1}{X \Lambda^2+m^2}$ 
 and for $D=4$ the space dimension in the $X$-variable is $d=2$ and $k=1$. We have then
\bwt
\begin{eqnarray}
\widetilde{\Big[\frac{1}{(p^2+m^2)}}\Big]_{\mu,D=4}&=&-\partial^2_{X}\big[\frac{X^2}{(X \Lambda^{2}+m^{2})}
\int^{\!\!\ ^{\mu^2  g(X)}}_{1}\!\!dt\frac{(1-t)}{t}\big].
\end{eqnarray}
\ewt
The full evaluation of Eq.(IV.8) is straightforeward. In keeping with the $D=2$ case \cite{PGEW06} the elimination of all non self-converging 
contributions in the $X$-integral requires that
 $g(x)=x^{(\alpha-1)}$ (up to a multiplicative arbitrary constant already taken into account 
as $\mu^2$) and that the limit $\alpha \rightarrow 1_{-}$ is performed. Then $h(x)=\mu^2 x^{\alpha}-1$  with $0< \alpha < 1$ is consistent with the
generic construction of $f(X)$ ({\it {cf}} Appendix B). It implies also $g(1)=1 > g(X_{max})=\frac{1}{\mu^2}$  {\it ie} $\mu^2>1$ and 
 $X_{max}=(\mu^2)^{(\frac{1}{(1-\alpha)})}$. In the limit $\alpha \rightarrow  1_{-}$ the upper integration limit in $X$  extends then to
  infinity, the $X-$integral is self-converging and the test function can be taken to unity over the whole integration domain. In this limit 
  the propagator at $x=y$ becomes
\bwt
\begin{eqnarray}
\Delta(0)&=&{\displaystyle \lim_{\alpha \rightarrow 1}}\int \frac{d^{4}p}{(2\pi)^4}\frac
{f^2(p^2)}{(p^2+m^2)}=2 m^4\int \frac{d^{4}p}{(2\pi)^4}\frac
{(\mu^2-1 -\log(\mu^2))}{(p^2+m^2)^3} \nonumber \\
&=&\frac{m^2}{(4 \pi)^2}(\mu^2 -1 - \log(\mu^2))               , 
\end{eqnarray}
\ewt
which is the expected result with respect to the scale parameter $\mu$ \cite{ITD}. One sees also that this parameter $\mu$ will
always be present independantly of the infinitely many possible partitions of unity ({\it cf} section II) that can be used to build up $f(X)$. In this sense $\mu$  is
universal, hence its relevance for the final RG-analysis. The fundamental achievement of Lagrange's formula is that it perfoms
the scaling analysis of the distribution $T(X)$ through the integral on the $t$ variable which incorporates the relevant scaling 
informations carried by $\mu$. The analysis of cases involving loops (two- and four-points vertex functions) follows the
same line and is detailed in Appendix C.\\
\par
 Technically things are a bit more involved in Minkowski's metric. The
test function becomes a function $\rho(x_{0},\overrightarrow{x})$
in coordinate space or $f(p_{0}^{2},\overrightarrow{p}^{2})$
in momentum space. As usual it is preferable to first carry out the
integrations over $x_{0}$ or $p_{0}$ respectively and then the remaining
integrals. There are two possible situations:\\
\par
$(1)$ The integrals over $x_{0}$ or $p_{0}$ are convergent. In this case
the remaining integrations are as in the Euclidean space with dimension
$D - 1$ and the extension is calculated correspondingly. As is shown
in section V there is no divergence problem for the $p_{0}$-integration
in the calculation of the retarded and advanced extended distributions
of causal distributions.\\
\par
$(2)$ The integrals over $x_{0}$ or $p_{0}$ diverge. This requires an
extension of the distribution with respect to the dependance on $x_{0}$ or
$p_{0}$.\\
\par
For practical reasons the calculations are usually done
in momentum space and we restrict the discussion to this
case. In principle there can be an UV-divergence problem of the $p_{0}$-integration 
or there can be nonintegrable poles at finite values
of $p_{0}$. The latter singularities are of IR-nature. Such situations
can arise when extending non-causal distributions like powers
of Feynman propagators. Once the $p_{0}$-extended distribution has
been calculated the remaining integrations are to be done as in $(1)$.\\
As a matter of illustration we recalculate the (Euclidean) result
of eq.(IV.9) in Minkowski's metric (example of integrable pole contribution).
We want to give a meaning to the diverging integral $\Delta(0)$ at $D=2,4$
\bwt
\begin{eqnarray}
\Delta(0)&=&\int d^Dp \frac{ f^2(p_0^2,p^2)}{p_0^2-p^2-m^2+2i \epsilon \omega_p} \nonumber \\
        &=&-\int \frac{d^{(D-1)}p}{2 \omega_p}\int^{\infty}_{-\infty}dp_0[\frac{1}{\omega_p-p_0-i\epsilon}+\frac{1}{\omega_p+p_0-i\epsilon}]
	f^2(p_0^2,p^2) 
\end{eqnarray}
\ewt
The $p_0$-integration cannot be done using contour integration because the extension of the test function to the whole complex plane
is not possible in general. However one can proceed as follows. One has
\bwt
\begin{eqnarray}
PP\int^{\infty}_{-\infty}dp_0\frac{f^2(p_0^2,p^2)}{p_0\pm\omega_p}&=&{\displaystyle \lim_{\epsilon \rightarrow 0}}
\{\int^{\mp \omega_p -
\epsilon}_{-\infty}dp_0+\int_{\mp \omega_p +\epsilon}^{\infty}dp_0\}\frac{1}{p_0\pm\omega_p}[-p_0\frac{d}{dp_0}\int_1^\infty\frac{dt}{t}
f^2(p_0^2t^2,p^2)] \nonumber \\
&=&{\displaystyle \lim_{\epsilon \rightarrow 0}}[\pm \frac{\omega_p}{\epsilon}-1 \mp \frac{\omega_p}{\epsilon} \pm  \frac{\omega_p}{\epsilon}
+1 \mp \frac{\omega_p}{\epsilon}]\log[\mu^2]=0,
\end{eqnarray}
\ewt
\par
where a partial integration on $p_0$ has been performed. Hence 
\begin{equation}
\int^{\infty}_{-\infty}dp_0\frac{f^2(p_0^2,p^2)}{p_0 \pm \omega_p \mp i\epsilon}=\pm i\pi f^2(\omega^2_p,p^2).
\end{equation}
At dimension $D=2$ one obtains
\begin{eqnarray}
& & \Delta(0)=-i\pi\int^{\infty}_{-\infty} \frac{dp}{\omega_p}f^2(\omega^2_p,p^2), \\
& &=-i\pi<\frac{1}{\omega_p},f^2(\omega^2_p,p^2)>=-i \pi
 <\widetilde{\frac{1}{\omega_p}},1>.\nonumber
\end{eqnarray}
Here $d=1$,\quad $\omega=0$, and  $k=0$. Thus
\begin{eqnarray}
\widetilde{(\frac{1}{\omega_p})}&=&\frac{d}{dp}[p \int_1^{\mu^2}\frac{dt}{t} \frac{1}{\sqrt{p^2+m^2t^2}}],\nonumber\\
&=&\frac{1}{\sqrt{p^2+m^2}}-\frac{1}{\sqrt{p^2+m^2\mu^4}}.
\end{eqnarray}
The end result for $\Delta(0)$ scales then as $\log[\mu^2]$ as expected. The two contributions in Eq.(IV.14) may be recombined to give 
a PV-type of substraction at the level of the propagator, as discussed more extensively in the next sub-section
\begin{eqnarray}
\Delta(0)&=&\int d^2p \frac{ f^2(p_0^2,p^2)}{p^2-m^2+i \epsilon}, \\
 &=&\int d^2p [\frac{1}{p^2-m^2+i\epsilon}-\frac{1}{p^2-m^2\mu^4+i\epsilon}]. \nonumber	 
\end{eqnarray}
It is important to note that causality is restored since the non-causal diverging term in $\delta(x^2)$ now present in each 
contribution of Eq.(IV.15) just cancels out in the substraction. Here this is a consequence of "repairing" an ill defined
 Feynman propagator. In CPT it is a general feature that - by construction - diverging causality
  violating terms are avoided right from the beginning \cite{Scharf}. At dimension $D=4$ one has 
\begin{eqnarray}
 \!\!\!\!\!\!\Delta(0)&=&-4i\pi^2 \Lambda^3\int^{\infty}_{0} \frac{X dX f^2(X)}{\sqrt{X(X \Lambda^2+m^2)}},\nonumber \\
\!\!\!\!\!\!&=&-4i\pi^2 \Lambda^3 <\widetilde{\frac{1}{\sqrt{X(X \Lambda^2+m^2)}}},1>.
\end{eqnarray} 
Here $d=2$,\quad $\omega=0$, and $k=1$. Thus
\bwt 
\begin{eqnarray}
\!\!\!\!\!\!& &\widetilde{(\frac{1}{\sqrt{X(X\Lambda^2+m^2)}})}=-\partial^2_{X}[X^2\int_1^{\mu^2} dt\frac{(1-t)}{t^2}\frac{1}{\sqrt{X(X\Lambda^2+m^2t)}}].
\end{eqnarray}  
\ewt    
The final $t$ and $X$ integrations give  back the result of Eq.(IV.9).
\subsection{Transcription of Epstein-Glaser method in terms of Pauli-Villars  type of substractions at the level of propagators}
Again we consider the scalar propagator $\Delta(x-y)$ in dimension $D=2$ for the simplest demonstration of this transcription.
It proceeds as follows.\\
An alternative form of $\widetilde{T}^{>}(X)$ is obtained through the change of variables $Xt \rightarrow Y$ in the first line of
 Eq.(IV.6). It gives
\begin{equation}
\widetilde{T}^{>}(X)=\frac{(-1)^k}{k!}X^{(k-d+1)}\partial^{(k+1)}_{X}\big[X^d \int^{\!\!\
^{\mu^2}}_{1}\!\!\!\!\!\!dt\frac{(1-t)^{k}}{t^{(d+1)}} T(\frac{X}{t})\big]
\end{equation}
\newpage
Setting $\|X\| \equiv  X$ as before, we have, in Euclidean metric  for $D=2$, $d=1$ and $k=0$ which gives
\begin{eqnarray}
\widetilde{\Big[\frac{1}{(p^2+m^2)}}\Big]^{alter}_{\mu,D=2}&=&\partial_{X}\big[X
\int^{\!\!\ ^{\mu^2}}_{1}\frac{dt}{t} \frac{1}{(X \Lambda^{2}+m^{2} t)}\big] \nonumber\\
&=&\frac{1}{(p^{2}+m^{2} )}-\frac{1}{(p^{2}+m^{2}\mu^2 )}.\nonumber
\end{eqnarray}
This is a Pauli-Villars subtraction, however without any additionnal scalar field. The final momentum integration gives 
the familiar RG-invariant result $\Delta(0)=\frac{1}{(4\pi)}\log(\mu^2)$.
The situation for $D=4$ is a bit more intricate but also carries usefull informations about the limitations of the PV-subtractions with respect
to the scaling analysis. As seen above the dimension in the $X$ variable is $d=2$ and $k=1$. The scalar propagator at $x=y$ takes then a form
somewhat different from that given in Eq.(IV.8),
\bwt 
\begin{eqnarray}
\widetilde{\Big[\frac{1}{(p^2+m^2)}}\Big]^{alter}_{\mu,D=4}&=&-\partial^{2}_{X}\big[X^2
\int^{\!\!\ ^{\mu^2}}_{1}\frac{dt}{t^2}(1-t) \frac{1}{(X \Lambda^{2}+m^{2} t)}\big] \\
&=&-2\int^{\!\!\ ^{\mu^2}}_{1}\frac{dt}{t^2}(1-t) \big[\frac{1}{X \Lambda^{2}+m^{2} t}-\frac{2 X \Lambda^2}{(X \Lambda^{2}+m^{2} t)^2}+
\frac{X^2 \Lambda^2}{(X \Lambda^{2}+m^{2} t)^3}\big].\nonumber
\end{eqnarray}
\ewt
It is verified that the final $X$-integration gives the very same result for $\Delta(0)$ as in Eq.(IV.9). In Eq.(IV.19) the expression 
$I(X,m)$ in brackets is regular at $X=0$  and behaves as $\frac{1}{X^3}$ for $X \rightarrow \infty$. 
Relation (IV.19) is a particular type of PV substraction. Indeed it may be written as (ignoring $t$ and $\Lambda $ which are inessential for the 
argument)
\begin{eqnarray}
I(X,m)&=&\frac{1}{X +m^{2}}\big [1-\frac{\alpha X+\beta}{X+\Lambda^2_1}+\frac{\gamma X^2+\delta X+\epsilon}{(X+\Lambda^2_1)(X+\Lambda^2_2)}
\big ] \nonumber \\
&=&\frac{A+B X+C X^2}{(X +m^{2})(X+\Lambda^2_1)(X+\Lambda^2_2)},
\end{eqnarray}
with $\alpha=2,\gamma=1,\beta=\delta=\epsilon=0$ and $\Lambda_1=\Lambda_2=m$. In the general form of Eq.(IV.20), imposing the fall off 
in $\frac{1}{X^3}$ for $X \rightarrow \infty$ gives $B=0,C=0$. Without loss of generality one may take $\delta=\epsilon=0$ then
$\beta=\Lambda^2_1+(1-\alpha)\Lambda^2_2$ , $\gamma=(\alpha-1)$ and $A=\Lambda^4_2(\alpha-1)$. Normalizing such that
$\frac{A}{(m^2-\Lambda^2_1)(m^2-\Lambda^2_2)}=1$ determines $\alpha$ to give the known result that a general regularisation of 
$\frac{1}{p^2+m^2}$  may proceed through multiplication by ${\displaystyle \Pi^n_{i=1}} \frac{(\Lambda_i^2-m^2)}{(p^2+\Lambda_i^2)}$ with the
decomposition in terms of PV contributions
\bwt
\begin{equation}
\frac{1}{p^2+m^2}{\displaystyle \Pi^n_{i=1}} \frac{(\Lambda_i^2-m^2)}{(p^2+\Lambda_i^2)}=\frac{1}{p^2+m^2}-{\displaystyle \sum^n_{j=1}}
\frac{{\displaystyle \Pi^n_{i=1\neq j}} (\Lambda_i^2-m^2)}{{\displaystyle \Pi^n_{i=1\neq j}} (\Lambda_i^2-\Lambda_j^2)}
\frac{1}{p^2+\Lambda_j^2}.
\end{equation}
\ewt
Under this form it is clear that each individual  PV-term treated separetly would have to undergo Epstein-Glaser treatment to exhibit
its scaling behaviour in terms of $\mu$. It shows that even though $F_\Lambda(p^2)={\displaystyle \Pi^n_{i=1}} \frac
{(\Lambda_i^2-m^2)}{(p^2+\Lambda_i^2)}$ might be viewed as a special case of test function reducing to unity when $\{\Lambda_i\} \rightarrow 
\infty$ and obeying Lagrange's formula Eq.(IV.1) with $X=p^2$  it does not have the intrinsic properties
of partition of unity test functions beeing essential for the  scaling analysis as described in the preceeding paragraph.
\setcounter{equation}{0}
\section{Causal splitting of distributions and link with dispertion relation technics}
Here the calculations of the retarded  extension of a singular distribution is worked out explicitely. The result shows an interesting
link to substraction technics known from dispersion relations.\\
The starting point is the following form of Lagrange's formula for SRTF's:
\begin{equation}
\phi(X)\! =\!
(\omega+1)\!\!\!\!\!{\displaystyle \sum_{\mid \beta \mid=\omega+1}}\!\!\!\big[
\frac{X^{\beta}}{\beta !}\!\!\int^{1}_{0} \!\!\!\!dt \frac{(1-t)^{\omega}}{t^{(\omega+1)}} \partial^{\beta}_{(X)}\phi(Xt)\big].
\end{equation}
It yields $\widetilde{T^<_r}(X)$ after partial integrations and taking into account the restriction brought about by $\phi(Xt)$ in the
$t-$integral ({\it cf} Appendix B, Eq.(B.3) and the discussion thereafter):
\bwt
\begin{equation}
\widetilde{T^<_r}(X)=(-1)^{(\omega+1)}{\displaystyle \sum_{\mid \beta \mid=\omega+1}} \partial^{\beta}_{X}
\big[\frac{X^\beta}{\beta !} \int_{1/\mu^2}^{1}dt\frac{(1-t)^{\omega}}{t^{(d+\omega+1)}}\theta(\frac{v.X}{t}) T(\frac{X}{t})\big].
\end{equation}
\ewt
It corresponds to Eq.(IV.3) but with a less refined lower bound on the $t-$integral. However, in keeping with the PV-type of substraction
({\it cf} Eqs.(IV.14-15)), we shall see below that it provides also the interpretation of the subtraction in the dispersion relations.
The corresponding Fourier transform -which becomes $UV-$regulated \cite{foote3}, hence its notation- is:
\bwt
\begin{equation}
\overline{\widetilde{T^>_r}}(p)=\frac{(\omega+1)}{(2\pi)^{d/2}}{\displaystyle \sum_{\mid \beta \mid=\omega+1}} 
\big[\frac{p^\beta}{\beta !} \int_{1/\mu^2}^{1}dt\frac{(1-t)^{\omega}}{t^{(\omega+1)}}\int dk  \overline{\theta_v}(k)
\partial^\beta_p \overline{T}(pt-k)\big].
\end{equation}
\ewt
Using Eq.(III.15) for $\overline{\theta_v}(k)$ this becomes:
\bwt
\begin{equation}
\overline{\widetilde{T^>_r}}(p_1^0)=\frac{i}{2 \pi}\frac{(p_1^0)^{\omega+1}}{\omega!}\int_{-\infty}^\infty \frac{dk_1^0}{k_1^0+i\epsilon}
\int_{{1/\mu^2}}^{1}dt (1-t)^\omega\partial^{(\omega+1)}_{(p_1^0t)}\overline{T}(p_1^0t-k_1^0,0,\cdots,0).
\end{equation}
\ewt
With the change of variables $p_1^0t-k_1^0=k_1^{\prime 0}$ and after $\omega+1$ partial integrations in $k_1^{\prime 0}$ one obtains:
\bwt
\begin{equation}
\overline{\widetilde{T^>_r}}(p_1^0)=\frac{i}{2 \pi}\frac{(p_1^0)^{\omega+1}}{\omega!}(1-\mu^2)^{(\omega+1)}\int_{-\infty}^{\infty} 
dk_1^{\prime 0}\frac{\overline{T}(k_1^{\prime 0},0,\cdots,0)}{(p_1^0-k_1^{\prime 0}\mu^2+i\epsilon\mu^2)^{(\omega+1)}(p_1^0-k_1^{\prime 0}+
i\epsilon)},
\end{equation}
\ewt
where the final $t-$integration has been performed to give
\bwt
\begin{equation}
\int_{{1/\mu^2}}^{1}dt \frac{(1-t)^\omega}{(p_1^0t-k_1^{\prime
0}+i\epsilon)^{(\omega+2)}}=\frac{(\mu^2-1)^{(\omega+1)}}{(\omega+1)(p_1^0-k_1^{\prime 0}+i\epsilon)}\frac{1}{(p_1^0-k_1^{\prime 0}\mu^2
+i\epsilon\mu^2)^{(\omega+1)}}.
\end{equation}
\ewt
Finally with the substitution $k_1^{\prime 0}=s p_1^0$ in Eq.(V.5) and returning to the general frame \cite{Scharf} we get:
\bwt
\begin{equation}
\overline{\widetilde{T^>_r}}(p)=\frac{i}{2 \pi}\frac{(\mu^2-1)^{(\omega+1)}}{(\mu^2)^{(\omega+1)}}\int_{-\infty}^\infty ds
\frac{\overline{T}(sp)}{(s-1/\mu^2-i\epsilon^\prime)^{(\omega+1)}(1-s+i\epsilon^\prime)}.
\end{equation}
\ewt 
The retarded, extended distribution $\overline{\widetilde{T^>_r}}(p)$ satisfies an unsubstracted  dispersion relation since
$\widetilde{T^<_r}(X)$ of Eq.(III.12) is a well-defined regular quantity. On the other hand the original retarded distribution
 $T^<_r=\theta T$ is singular. A dispersion relation for this quantity would require $\omega+1$ substractions. It is interesting to see 
 that the explicit construction of the extended distribution in terms of the original singular distribution leads to the factor 
 $(s-1/\mu^2-i\epsilon^\prime)^{(\omega+1)}$ in the denominator of Eq.(V.7) which is characteristic of dispersion relations 
 with $\omega+1$ substractions, with one important difference: the substraction point is not arbitrary but $\frac{p}{\mu^2}$, as shown
  hereafter. It is  the scale $\mu$ present in the SRTF which fixes this point.
The calculation of the advanced, extended distribution $\overline{\widetilde{T^>_a}}(p)$ follows the same lines with a result similar to
 that of Eq.(V.7) with $i\epsilon \rightarrow -i\epsilon$. The difference of the retarded and advanced  pieces reduces to
\vspace{-1.4cm}
\bwt
\begin{equation}
\overline{\widetilde{T^>_r}}(p)-\overline{\widetilde{T^>_a}}(p)=\overline{T}(p)-{\displaystyle
\sum_{n=0}^{\omega}}\frac{(-1)^n}{n!}\big[\frac{\mu^2-1}{\mu^2}\big]^n \int_{-\infty}^{\infty}\!\!\!ds  \overline{T}(sp)  \delta^{(n)}(s-\frac{1}{\mu^2}
),\nonumber
\end{equation}
\begin{equation}
 \hspace{1.5cm}=\overline{T}(p)-{\displaystyle \sum_{n=0}^{\omega}}{\displaystyle \sum_{\mid \beta 
 \mid=n}}\frac{[\frac{p(\mu^2-1)}{\mu^2}]^\beta}{\beta!}\overline{T}^{(\beta)}(\frac{p}{\mu^2}), 
\end{equation}
\ewt 
which is just the Taylor remainder of $\overline{\widetilde{T}}(p-q)$ at $q=\frac{p}{\mu^2}$ -($\overline{\widetilde{T}}_q(p)$ in the notation
of \cite{Scharf})- that is the BPHZ substraction  at $q=\frac{p}{\mu^2}$ discussed in section III-E.\\

\setcounter{equation}{0}
\section{Non-perturbative Light-Front QFT (LFQFT) and the BSEG procedure} 
At various places in the text we commented on the necessity of using test functions when quantizing fields in Dirac's front
form \cite{Dirff}. As stated in the introduction the restriction of quantum fields to a lightlike surface does not canonically exist and
turns out to be the major mathematical problem to face first in dealing with LFQFT. The question was initially adressed in \cite{GUW}
and  fully discussed in \cite{ULL}. The outcome is the Fock expansion of the field operator written in LC-variables with test functions
taking proper account of both the LC-induced IR singularity at $k^+=0$ and of the UV behaviour in $k^+$. For a scalar field theory the
Haag series is well defined in terms of products of free fields and provides a non-perturbative LF-scheme to evaluate physical observables
by projection of the equations of motion and constraints on different Fock sectors \cite{GSW}. The procedure leads to coupled sets of
integral equations for amplitudes in the Haag series, presenting  non-standard renormalization difficulties common to all LF dynamics
approaches \cite{LFDMath,JHEM}. The main difficulty comes from the unavoidable truncation of the coupled set of equations to a finite number
of amplitudes. This truncation jeopardizes the renormalization procedure unless special attention is given to the scaling behaviour of the 
neglected contributions. It appears then that, if the BSEG procedure could be applied in the non-perturbative LFQFT context, then only finite
quantities will occur in the course of calculations and the amplitudes, instead of obeying coupled set of equations, could be determined
inductively. Thereby  overlapping contributions and the renormalization difficulties mentioned above could be avoided. An immediate objection
is that since the BSEG procedure is perturbative in essence it cannot fulfil our non-perturbative goals. However the same objection is 
also valid for the standard scheme with coupled equations and it is well known that the non-perturbative aspects of LFQFT are encoded in
the constrained  dynamics related to the singular nature of the light front Lagrangian. To be specific consider the standard expression for 
the S-matrix \cite{Kady}
\bwt
\begin{equation}
S=1+{\displaystyle \sum_{n}}\int (-i)^n H^{int}(x_1)\theta(t_1-t_2)H^{int}(x_2)\cdots \theta(t_{n-1}-t_n)H^{int}(x_n)d^4x_1\cdots d^4x_n,
\end{equation}
\ewt
where $H^{int}(x)$ is the interaction Hamiltonian  and the time ordering is made with $\theta$-functions, albeit the  divergences occuring on the
light cone when $(x_{i-1}-x_i)^2=0$ \cite{Carbo} and subjects of Epstein-Glaser attention. The light front is defined by $\omega.x=0$, 
where $\omega$ is an arbitrary four vector such that $\omega^2=0$. The expression of Eq.(VI.1) is then repesented in terms of the light
front time $\sigma=\omega.x$ for, if $(x_{i-1}-x_i)^2 > 0$, the signs of $\omega.(x_{i-1}-x_i)$ and $(t_{i-1}-t_i)$ are the same provided 
the time ordering is treated according to the generalized Epstein-Glaser considerations developped above which eliminates light-cone 
divergences. Hence $H_\omega^{int}=H^{int}$, however with important caveats. We discuss first the case of interacting scalar fields. 
Obviously the asymptotic states have to be free bosonic fields $\varphi_{0}(x)$ as defined after Eq.(II.4). Since the LC-Lagrangian is 
singular, the quantization has to follow the specific rules for constrained systems \cite{Di50,Ja93}~. In the present case it means that 
in addition to the equation of motion for the particle sector field $\varphi(x)$ there exists a constraint which lives only in the zero mode 
sector and is of nondynamical nature. If this constraint has a nontrivial solution $\Omega$ then the total field $\Phi(x)$ is a sum: 
$\Phi(x)= \varphi(x)+\Omega$. The zero mode field carries the information on nontrivial physics and is the LC-counterpart of a nontrivial 
vacuum in equal-time quantization. Consequently the field appearing in the interaction $H^{int}(x)$ is the total field $\Phi(x).$  There is,
however, a problem which cannot be overlooked: the zero mode $\Omega$ following from the constraint can only be obtained approximately and 
recursively in an iterative procedure, which means that the interaction $H^{int}(x)$ is a priori known only as far as $\Omega$ is known. 
This was the bad news, shared with alternative formulations such as that of Ref.\cite{JHEM} or the covariant light front dynamics approach 
of Ref.\cite{LFDMath}; the good news is that, if the coupling is large enough to support a zero mode - or in other words to generate a phase
transition - already the first, algebraically very simple iteration step leads to a form for the zero mode giving a value
for the critical coupling in good agreement with conventional, much more complex calculations and permitting to discuss the critical
 behaviour near the phase transition \cite{GSW}. Thus there is a good reason to hope that a small number of iteration steps would lead 
 to a proper description of non-perturbative physics.
The presence of zero modes in the field operator does not spoil the causality issues of the theory: causality expressed through the 
Pauli-Jordan commutator function makes a statement about the measurability of fields at different space-time points. Since such a statement 
concerns only the particle sector of the theory, the zero mode contribution is not involved in the causal commutator.\\
In fermionic theories the particle sector itself is split into a sum of independent and dependent degrees of freedom which leads 
-in gauge theories- to additional interactions. The zero mode problem concerns the gauge field itself and interaction terms built out of primary fermionic fields 
and behaving like scalar fields. \\
The validation of these ideas in LFQFT gauge theories are undoubtly challenging. Already Yang-Mills theories have been investigated in some 
details within the standard CPT framework \cite{Dutsch}, paving the way to the implementation of the CPT-LFQFT studies. They will be of 
considerable theoretical and even practical interests since the outcome is a rigourous non-perturbative alternative to lattice calculations,
enjoying the numerous benefits of an LFQFT formulation  put foreward for decades \cite{Brod05}.  
\section{Conclusions}
Our concerns in this paper were twofold. First to define in a mathematically
consistent way Quantum Fields as Operator Valued Distributions, focussing on
the specific test functions needed to achieve a generic QFT description 
preserving basic Poincar\'e and Lorentz invariances. In keeping with the general
analysis of integrals of differential forms the necessary test functions take
the form of partition of unity constructed from super regular building blocks on
compact subspaces. The second goal, related to the analysis of divergences
crippling the usual pertubative QFT approach, was to show, following 
Epstein and Glaser analysis, the role of a proper treatment of causality in 
extensions of singular distributions. These extensions can be performed very 
elegantly with Lagrange's integral kernel representation  combined 
with super regular test functions, for which the Lagrange
remainder of any order is identical to the test function itself. The
flexibility in the construction of the test functions allows to
treat IR- and UV-divergent distributions with the same mathematical
techniques. By the very nature of the construction an arbitrary scale,
characterising the building blocks of the decomposition of unity, comes
naturally into the picture. Such a scale is the corner stone permitting
the use of extended distributions in renormalization
group studies. Another important aspect of the decomposition of unity
concept for the construction of the test functions is that it does not violate
gauge invariance - contrary to other test function methods. A  first
treatment \cite{PGEW05} in the QED context verifies this for the polarisation
tensor in dimension $D=2$ and gives Fujikawa's analysis of the axial anomaly 
in dimensions $D=2,4$ directly from the presence of the test function as 
partiton of unity. 
A general analysis of gauge invariance is given in
\cite{Scharf} and should be further developped in the present context.
In coordinate space the extended distribution differs from the original
singular distribution only on the support of the singularity whereas in momentum
space it differs from the original one by subtractions
at points determined by the intrinsic scale of the test function.\\
Finally we presented arguments, based on the recent developments of Epstein-Glaser causal approach, which make it extremely plausible 
that a finite symmetry-preserving LCQFT could be envisaged on the basis of an iterative construction of the $S$-matrix
and a causality conditioned finite regularization using the OPVD treatments of fields.

\paragraph*{Acknowledgements}
E. Werner is grateful to Alain Falvard for his kind hospitality at LPTA and to the  IN2P3-Department from CNRS for financial support.
 We are particularly indebted to M. Slupinski for his expertise in topological analysis. 
Discussions with M. Weinstein, V. Braun, M. Lavelle, L. Martinovic and P. Ullrich  are gratefully acknowledged.

\vspace{0.5cm}
\appendix
\Large
\centerline{\bf Appendix }
\vspace{0.5cm}
\section{Composition of the product of two partitions of unity}
\small\normalsize
The purpose here is to give a proof of the Remark of  section II, quoted
quite frequently in the literature. Most of the time the proof is left to the
reader but since the property is basic to section II it is useful to establish
its validity.\\ 
\par
Property: If  $(\alpha_i)_{i\in I}$  and $(\beta_j)_{j\in J}$ are  two
partitions of unity subordinate to
$(U_k)_{k\in K}$  and $(V_l)_{l\in L}$ respectively,
$(\alpha_i\beta_j)_{(i,j)\in I\times J}$ is a  partition of unity
subordinate to $(U_k\cap V_l)_{(i,j)\in K\times L}$.\\

1) Clearly $(U_k\cap V_l)_{(i,j)\in K\times L}$ is a covering of ${\mathcal M}$\\

2) For $X\in {\mathcal M}$ one chooses a neighbourhood $U_X$ (resp. $V_X$) of $X$ such
all but a finite number of $\alpha_i$ (resp. $\beta_j$) are zero
on $U_X$ (resp. $V_X$). Then all but a finite number of the products
 $\alpha_i\beta_j$ are zero on the neighbourhood 
$U_X \cap V_X$ de $X$. This prooves condition d(i) of  section II  for the family of  functions
$(\alpha_i\beta_j)_{(i,j)\in I\times J}$.\\

3) Let $\alpha_1,\dots,\alpha_A$ et $\beta_1,\dots,\beta_B$ be the
$\alpha_i$ (resp. $\beta_j$ which does not vanish on $U_X$ (resp.
$V_X$). One has then for all 
 $x\in U_X$,
$$
\alpha_1(x)+\dots \alpha_A(x)=1
$$
and for all  $x\in V_X$,
$$
\beta_1(x)+\dots \beta_A(x)=1
$$
which implies that for all $x\in U_X\cap V_X$,
$$
(\alpha_1(x)+\dots \alpha_A(x)) (\beta_1(x)+\dots \beta_A(x))=1.
$$
The products $\alpha_i \beta_j$ appearing in this equation
are the only one which does not vanish on $U_X\cap V_X$, which implies that for
all $x\in U_X\cap V_X$, one has
$$
\sum_{(i,j)\in I\times J}\alpha_i(x)\beta_j(x)=1.
$$
This proves condition d(ii)of section II for the family of functions
$(\alpha_i\beta_j)_{(i,j)\in I\times J}$.\\

4) Since the support of a product of fonction is the intersection of the 
supports, if for  $i\in I$ on chooses $f(i)\in K$ such that
$supp \alpha_i\subseteq U_{f(i)}$ and for  $j\in J$ on chooses $g(j)\in L$ such that
$supp \beta_j\subseteq V_{g(j)}$ it will hold that
$supp(\alpha_i\beta_j) \subseteq U_{f(i)}\cap V_{g(j)}$.

This proves condition d(iii)of section II for the family of functions 
$(\alpha_i\beta_j)_{(i,j)\in I\times J}$.\\
\par
As a consequence the definition of the field given in Eq.(II-5) with $f$ as a partition of unity is independant of the way it is constructed
since $$\sum_{i\in I} \gamma_i\equiv \sum_{(i,j)\in I\times J}\alpha_i\beta_j=\sum_{(j,i)\in J\times I}\beta_j\alpha_i\equiv \sum_{j\in J} \nu_j
$$. 
\section{Test functions as partition of unity: construction and properties.}
\small\normalsize
 Since in the analysis of Section IV the generic test function depends only on one variable, the
dimensionless variable $\| X \| $,  here we only  need to detail  the simplest case of the decomposition of unity on the line.
The family of functions $\{\beta_{j}\}$  building up unity are based on an elementary function $u$ with the 
 property that for any point $x \in [0,h]$ \quad \!\!\!$u(x)+u(h-x)=1$. This is the simplest realisation of the
 local finiteness property of the $\{u_j\}$ family. By translation for any $j \in I$  and for any $x \in [jh,(j+1)h]$ one has 
$u(x-jh)+u((j+1)h-x))=1$. Then $\beta_{j}(x)=u(x-jh)$ and $\sum_{j=-\infty}^{\infty}\beta_{j}(x)=1$. The overlaping 
subsets $(O_{j})$ covering the line are visualised in Fig.$(B1)$.
\bwt
\begin{center}
\vspace*{-4cm}
\epsfig{file=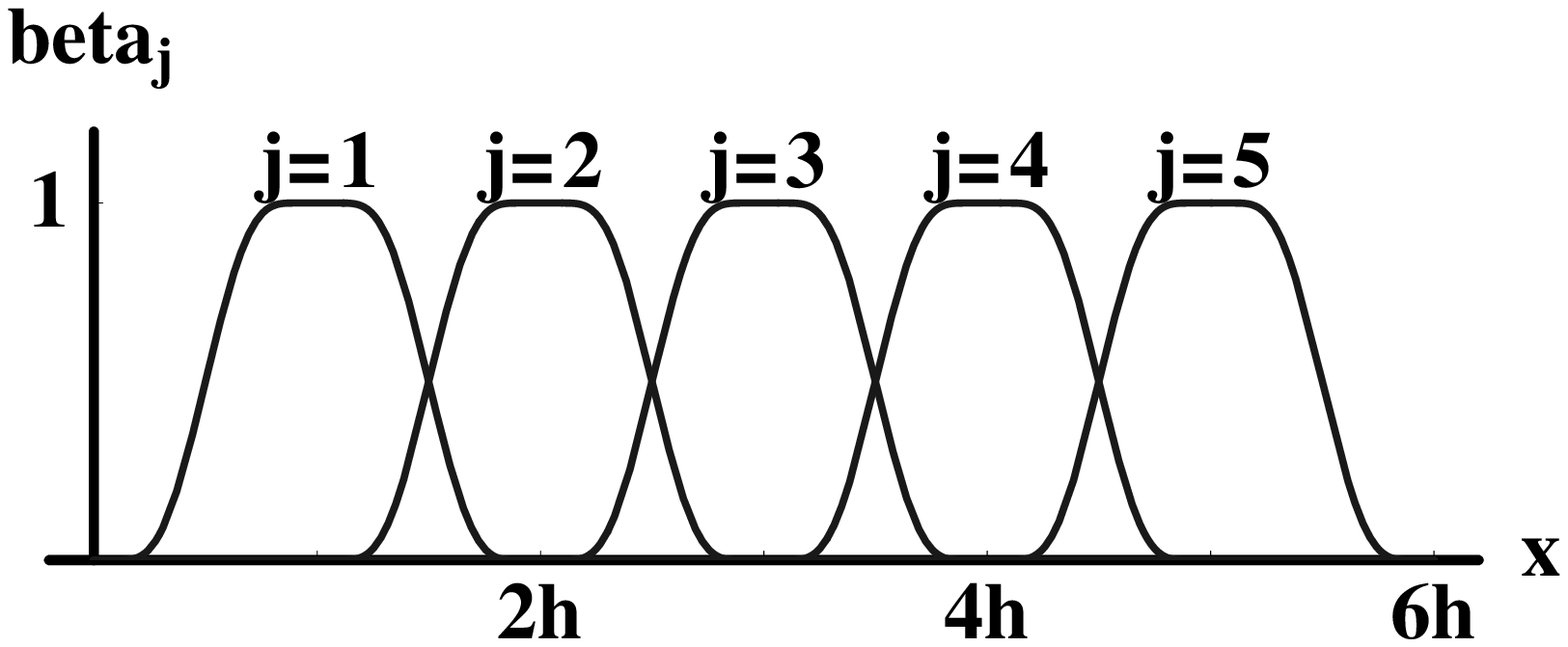,height=12cm,width=12cm}\\
\vspace*{-4cm}
Fig.B1: The functions $\beta_j(x)=u(x-jh)$ decomposing unity.
\end{center}
\ewt
 Each individual $\beta_{j}$ vanishes with all its derivatives outside $(O_{j})$.
 Due to the independence of the OPVD functional on the construction of the partition of unity there is no need to consider more 
 elaborate constructions in terms of the constituants functions $\{u_j\}$. From the relation $u(x-jh)+u((j+1)h-x))=1$ for 
  $x \in [jh,(j+1)h]$  it follows also that 
 \begin{eqnarray} 
 \gamma_j(k)&=&\int_{(j-1)h}^{(j+1)h}\quad \!\!\! dx\quad \!\!\! u(x-jh)\quad \!\!\! e^{-ikx} \\ 
            &=& 0\quad \mathrm{if}\quad k=\frac{2 n \pi}{h},\quad h\quad \mathrm{if}\quad k=0,\quad \forall j.\nonumber
 \end{eqnarray} 
 This orthogonality property leads to the Fourier mode analysis in the distributional context. It is also interesting to note that the 
 orthogonal Bloch functions approach used in the noncompact lattice formulation of gauge theories of Ref.\cite{Freid} can be then rephrased 
 in terms of the decompositon of unity, the lump functions there being particular realisations of the $u(x-jh)$ with only a finite
 number of derivatives vanishing outside $(O_{j})$. For completness we give explicitely two constructions of the elementary function
 $u$. They are based on the regularising function of L. Schwartz \cite{LSCW}
\begin{equation} 
\rho(x)={\mathcal N} e^{\frac{1}{x^2-1}} \quad \mathrm{if} \quad |x| < 1, \quad 0 \quad \mathrm{if} \quad |x| \geq  1,
\end{equation}
where ${\mathcal N}$ stems from normalising $\rho(x)$ to unity. $u^{(j,h)}_{\epsilon}(x)$ is then obtained from the convolution of the
characteristic function $\{\chi_{j,h}(x)=1 \quad \mathrm{if} \quad jh \leq x \leq (j+1)h,\quad 0$ elsewhere $\}$,  with the function  
$\rho_\epsilon(x)$ such that $\rho_\epsilon(x)=\frac{1}{\epsilon}\rho(\frac{x}{\epsilon})$. One has $u^{(j,h)}_{\epsilon}(x) \neq 0\quad
\mathrm{for}\quad x \in [jh -\epsilon,(j+1)h+\epsilon]$ and $u^{(j,h)}_{\epsilon}(x) =1 \quad
\mathrm{for}\quad x \in [jh +\epsilon,(j+1)h-\epsilon]$. Then $\sum_{j=0}^{N-1} u^{(j,h)}_{\epsilon}(x)$ is a partition of unity
 on $\Omega=[0,L]$ with $L=Nh$. Another useful construction is achieved, for any $\nu >0$, with
\begin{equation} 
u(x-jh)= \left\{
\begin{array}{lll}
& &\!\!\!\!\!\!{\mathcal N}\int_{|x-jh|}^{h} dv \exp[-\frac{h^{2\nu}}{v^{\nu} (h-v)^\nu}],\\
& &\quad \mathrm{for} \quad (j-1)h < x <(j+1)h; \\ \\
&0& \quad \mathrm{for} \quad |x-jh|\geq h,
\end{array}
\right.   
\end{equation}
where ${\mathcal N}=\int_0^h dv \exp[-\frac{h^{2 \nu}}{v^{\nu} (h-v)^\nu}]$. It is verified that for any $x \in [jh,(j+1)h]$ one has indeed 
$u(x-jh)+u((j+1)h-x))=1$ and that $\beta_{j}$ vanishes with all 
its derivatives outside $(O_{j})=[(j-1)h,(j+1)h]$.\\
In the scaled variable $\| X \| $ the final test function $f(X) \equiv f^{>}(X) $ built in this way is then 
{\bf
\begin{equation} 
f(X)\equiv f^>(X)= \left\{
\begin{array}{lll}
&\mathrm{(1-\chi(\| X \|+1,h))}& \mathrm{for~} 0 < \| X \| \leq h\\
&1&              \mathrm{for~} h < \| X \| \leq 1  \\
&\mathrm{\chi(\| X \|,h)}&\mathrm{for~} 1 < \| X \| < 1+h  \\
&0&  \mathrm{for~} \| X \|  \geq 1+h 
\end{array}
\right.   
\end{equation}}
\bwt
\begin{center}
\vspace*{-9cm}
\epsfig{file=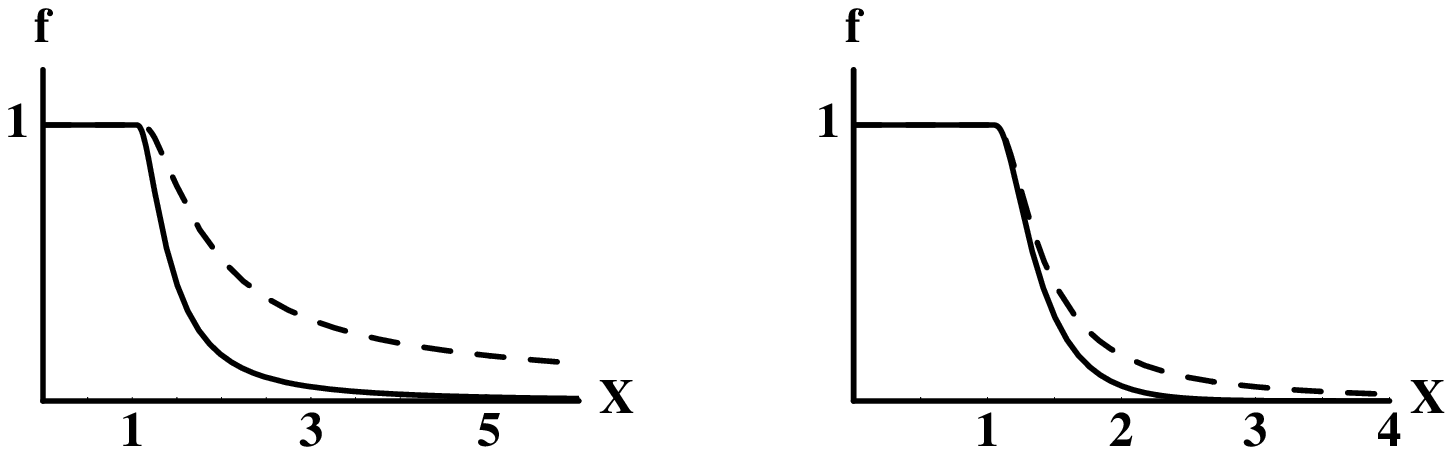}\\
\vspace*{-9cm}
 Fig.B2: The function $f(X)$ from Eqs.(B4) and from $u(x-jh)$ with $\nu=1$\\
\hspace*{1cm} a)left curves $\alpha=0.95$, dashed: $\mu^2=1.25$; full line: $\mu^2=1.15$;\\
\hspace*{1.2cm} b)right curves $\mu^2=1.15$, dashed: $\alpha=0.95$; full line: $\alpha=0.85$;      
\end{center}
\ewt
Hence the function  $\chi(\| X \|,h)$ is just the very end  part of the elementary function $u$ building up unity, 
say $u( \| X \| -1)$ in the last example.
Since $f(X)$ vanishes with all its derivatives at $\| X \|=1+h$, Lagrange's formula, Eq.(IV.5), is identically fulfilled. Moreover
from its construction  $f(Xt)$ is different from zero $\mathrm{if} \quad \| X \|t < (1+h)$. Hence the upper limit of the t-integral 
in Eq.(IV.5) is $t_{max}=\frac{(1+h)}{ \| X \|}$. Due to the regularising exponential  at the upper boundary  it is easily seen on the 
two constructions presented above that $f(X)$  still vanishes  at an $\| X \|$-dependant upper boundary with all its derivatives and
faster than any inverse power of $\| X \|$. This property is assumed to be preserved for the generic function $\chi(\| X \|,h(\| X \|))$. 
One writes now $h(X)=\mu^2 \| X \|g( \| X \|)-1$ where $\mu$ is an arbitrary dimensionless parameter extracted from $g(\| X \|)$. 
It controls then the size of the overlapping regions for two consecutives $(O_{j})$.This  is shown in 
Fig.$(B2)$,where $f(X)$, constructed from Eqs.(B4) and from $u(x-jh)$ with $h(X)=\mu^2 \| X \|^{\alpha}-1$, is shown respectively at 
fixed $\alpha ({\it viz.} \mu^2)$ for two values of $\mu^2 ({\it viz.} \alpha)$.  With the scale in the X variable used in Fig.$(B2)$ 
for $0 < \| X \| \leq h$ \quad $f(X)$ rises so sharply to 1 that it is not distinguishable from a $\theta-$function.\\  
 From the discussion in Section IV \quad $\mu > 1$ and $t_{max}=\frac{(1+h)}{ \| X \|}=\mu^2 \| X \|^{(\alpha-1)} > 1$. The maximum 
 value of $\| X \|$ is $(\mu^2)^{(\frac{1}{1-\alpha})}$ and tends to infinity when $\alpha$ tends to $1$.
 For a distribution $T(X)$ singular at the origin of $\RR^{d}$ and with the representation of Eq.(B.3) the test function $f^{<}(X) $  behaves
 simply  as  $w(X)=u(h-\|X\|)\quad \mathrm{for~} \quad \| X \| \leq h $. Then in the
 limit $h \rightarrow 0$  $w(X)$  tends to a step function $\theta(\|X\|-h)$. When $\alpha \rightarrow 1$ \quad $h$ effectively tends to zero
  for values of $\|X\|$ of the order of $\frac{1}{\mu^2}$ and then $w(\frac{X}{t})=\theta(\frac{\|X\|}{t}-h(\frac{\|X\|}{t}))=
 \theta(t-\|X\|(\mu^2-1))$. Since the extension of $T(X)$ is seeked in the IR and $\tilde{\mu}=(\mu^2-1) > 0$, $\tilde{\mu}\|X\| < 1$  is a
  lower bound to the $t$-integral in Eq.(IV.3).\\

\setcounter{equation}{0}
\Large
\section{ Evaluation of loop-contributions with partition of unity test functions.}
\small\normalsize
\par 
We consider first the one-loop contribution $I(k^2)$ to the four-point function of $\Phi^4$ scalar field theory at $D=4$. We shall work in 
Euclidean metric to simplify the derivation. $I(k^2)$ reads
\begin{eqnarray}
I(k^2)\!\!&=&\!\!\int\!\!\frac{d^4p}{(2 \pi)^4} \frac{f^2(p^2)f^2((p+k)^2)}{(p^2+m^2)((p+k)^2+m^2)},\\
     \!\!&=&\!\!\int^{1}_{0}\!\!dx \int \!\!\frac{d^4q}{(2 \pi)^4}\frac{f^2((q+x k)^2))f^2((q-(1-x)k)^2)}{(q^2+k^2 x(1-x)+m^2)^2}.\nonumber
\end{eqnarray}
The two cases $k^2=0$ and $k^2\neq 0$ must be distinguished. In the first case, with $X=\frac{p^2}{\Lambda^2}$, one has
\begin{equation}
I(0)=\frac{\Lambda^4}{(4 \pi)^2} \int_{0}^{\infty} \frac{X dX f^4(X)}{(X \Lambda^2+m^2)^2}.
\end{equation}
With $T(X)=\frac{X}{(X \Lambda^2+m^2)^2}$, ${\displaystyle \lim_{X \rightarrow \infty}} T(X) \sim \frac{1}{X}$. Hence $d=1, \omega=0, k=0$
and then
\begin{equation}
\widetilde{T}(X)=\partial_{X}[\frac{X^2}{(X \Lambda^2+m^2)^2}]=\frac{2 X m^2}{(X \Lambda^2+m^2)^3}
\end{equation}
It follows that 
\begin{equation}
I(0)=\frac{2 m^2 \Lambda^4}{(4 \pi)^2} \int_{0}^{\infty} \!\!\!\!\frac{X dX}{(X \Lambda^2+m^2)^3}\int_{1}^{\mu^2}\!\!\frac{dt}{t}=\frac{1}{(4 \pi)^2}
\log(\mu^2)
\end{equation}
In the second case, $ k^2\neq 0$, since the test function is unity for small values of its argument, in the UV regime $x\| k\|$ and $(1-x)\|k\|$ can be
 disregarded with respect to $\|q\|$ and the regulation of $I(k^2)$ is provided by $f^4(q^2)$ only (This can be checked explicitely by Taylor expansion of the
$f$'s, for in all explicitely finite contributions the limit $\alpha \rightarrow 1$ can be taken, then $f \rightarrow 1$ everywhere and all 
 terms involving derivatives of $f$ with respect to $q^2$ do not contribute). Eq.(C.1) transforms to
\begin{eqnarray}
I(k^2)\!\!&=&\!\!\frac {\Lambda^4}{(4 \pi)^2 m^4}\int^{1}_{0}\!\!\!dx \int_{0}^{X_{max}}\!\!\!\!\!\!\!\!\!\!\!\!dX\frac{X f^4(X)}{(X \frac{\Lambda^2}{m^2}+ \frac{k^2}{m^2}x(1-x)
+1)^2},\nonumber \\
\!\!&=&\!\!\frac {\Lambda^4}{(4 \pi)^2 m^4}\int^{1}_{0}\!\!\!dx \int_{0}^{Y_{max}}\!\!\!\!\!\!\!\!\!\!dY\frac{Y f^4[Y(\frac{k^2}{m^2}x(1-x)
+1)]}{(Y \frac{\Lambda^2}{m^2}+1)^2}\nonumber\\
\hspace*{-1cm}& &
\end{eqnarray}
where, in the last equation, $X$ has been changed to $Y(\frac{k^2}{m^2}x(1-x)+1)$. As before \\
$d=1, \omega=0, k=0$ and $I(k^2)$ reads now
\begin{eqnarray}
I(k^2)&=&\frac{2 \Lambda^4}{(4 \pi)^2 m^4} \int^{1}_{0}dx\int_{0}^{Y_{max}}\frac{Y dY}{(Y\frac{\Lambda^2}{m^2}+1)^3}\nonumber\\
& &\int_{1}^{\infty}\frac{dt}{t}f^4[Y t(\frac{k^2}{m^2}x(1-x)+1)]
\end{eqnarray}
As discussed in Appendix B the test function effectively cuts the t-integration such that $t \leq \frac{\mu^2}{(\frac{k^2}{m^2}x(1-x)+1)}$ 
in the limit $\alpha \rightarrow 1$, where $Y_{max}\rightarrow \infty$ and  the test function tends to unity over the whole 
integration domain in $Y$. Hence 
\begin{eqnarray}
I(k^2)&=&\frac{1}{(4 \pi)^2 } \int^{1}_{0}dx\int_{1}^{\frac{\mu^2}{(\frac{k^2}{m^2}x(1-x)+1)}}\frac{dt}{t} \\
      &=&\frac{1}{(4 \pi)^2 }\{\log(\mu^2)-\int^{1}_{0}dx\log[\frac{k^2}{m^2}x(1-x)+1]\}.\nonumber
\end{eqnarray}
>From the analysis in dimension $D=4-\epsilon$ the result is known to be \cite{ITZ}
\begin{eqnarray}
I(k^2)&=&\frac{\Gamma(\epsilon/2)}{(4 \pi)^2 } m^{-\epsilon/2} \int^{1}_{0}\!\!\!\!dx [\frac{k^2}{m^2}x(1-x)+1]^{-\epsilon/2}  \\
      &=&\frac{1}{(4 \pi)^2 }\{\frac{2}{\epsilon}-\int^{1}_{0}\!\!\!\!dx\log[\frac{k^2}{m^2}x(1-x)+1]\} + Cte. \nonumber
\end{eqnarray} 
As expected this is another example that the infinite contributions in $\frac{2}{\epsilon}$  present in the conventionnal formulation has been
removed and replaced by the scale dependance originating from the overlapping domains building up the partition of unity. In a 
regularisation procedure  by a cut-off the divergence is $ \log(\frac{\Lambda^2}{m^2})$ and, according to Eq.(IV.15), the regulator mass is 
$\Lambda=\mu m $, hence the final result in $\log(\mu^2)$.\\
 Instead of the derivation retained here one may use as well Schwinger's representation for the propagators. The test functions in this case just
provide the necessary handling of divergences of the final integrals over Schwinger's parameter, with the very same result of Eq.(C.7).The
way the regularisation works is most simply seen for $\Delta(0)$ at $D=4$. One has
\begin{eqnarray}
\Delta(0)&=&\int \frac{d^4q}{(2\pi)^4}\frac{f(q^2)}{q^2+m^2} \\          
         &=&\frac{\Lambda^4}{(4\pi)^2 m^2} \int_0^\infty y dy e^{-y\frac{\Lambda^2}{m^2}}\int_0^\infty \frac{du}{u^2} e^{-u}
	 f(\frac{y}{u}).\nonumber\\
& & \nonumber	 
\end{eqnarray}	
 
Now for a SRTF it holds that
\begin{eqnarray}
f(\frac{y}{u})&=&-\int_1^\infty dt(1-t)\partial_t^2 \big[t f(\frac{yt}{u})\big] \nonumber\\
&=&-u^2\int_1^\infty \frac{dt}{t} (1-t) \partial_u^2 f(\frac{yt}{u}).
\end{eqnarray}
Hence the test function takes care of the divergence at $u=0$, and after partial integrations all integrals are now finite giving the
result of Eq. (IV.9).\\
We examplify now the general handling of loop contributions with Schwinger's parametrization for the two-point function $\Gamma_2[{\bf k}]$ with
two loops. It reads
\bwt
\begin{eqnarray}
\Gamma_2[{\bf k}]&=&\int \frac{d^4q_1d^4q_2 }{(2 \pi)^8} \frac{f^2(q_1^2)f^2(q_2^2)f^2((q_1+q_2-k)^2)}{(q_1^2+m^2)(q_2^2+m^2)((q_1+q_2-k)^2+m^2)} \nonumber \\
		 &=&\int  \frac{d^4q_1d^4q_2 d^4q_3}{(2 \pi)^8} \frac{\delta(q_3+k-q_1-q_2)f^2(q_1^2)f^2(q_2^2)f^2(q_3^2)}{(q_1^2+m^2)
		 (q_2^2+m^2)(q_3^2+m^2)} \nonumber \\
		 &=& \int d^4y e^{i k.y}\prod_{i=1}^{3}  \int  \frac{d^4q_i}{(2 \pi)^4} e^{ -i q_i.y}  \frac{f^2(q_i^2)}{(q_i^2+m^2)}
		 \nonumber \\
		 &=& \int d^4y e^{i k.y}\prod_{i=1}^{3} \int d\alpha_i e^{-\alpha_i m^2}\int  \frac{ d^4q_i}{(2 \pi)^4}
e^{ -\alpha_i q_i^2-i q_i.y} f^2(q_i^2).
\end{eqnarray} 
\ewt
In the last relation  $f^2(q_i^2)$ depends in fact only upon the argument $X_i=\frac{q_i^2}{\Lambda^2}$. Then with the change of integration
variable $\alpha_i q_i^2 \rightarrow \widetilde{q_i}^2$  the argument of the test function becomes $\frac{\widetilde{q_i}^2}{\alpha_i \Lambda^2}$
which implies ({\it cf} Appendix B) $\alpha_i \Lambda^2 > \frac{1}{\mu^2}.$  The integrals over $\alpha_i$ have a lower boundary cut-off at
 $\frac{1}{\mu^2 \Lambda^2}$, just as in the conventionnal treatment \cite{ITD}. However using instead Lagrange's 
formula for the test functions gives a finite result, whose dependence on the scale parameter $\mu$ can be inferred from the conventionnal analysis with the
replacement  $\Lambda \rightarrow \mu m $.
\vskip5pt
\setcounter{equation}{0}
\Large
\vspace{0.5cm}
\section{Recollection of Lagrange's types of integral formulae for super-regular test functions.}
\small\normalsize
\vspace{0.5cm}
 In this section we collect integral formulae used  alternatively in the UV and IR regimes at different places in the main text and 
 show their equivalence.\\
  The first one encountered in the IR is given in Eq.(IV.1). At $D=1$, after $k$ partial integrations, the following identity also holds for
  SRTF 
$$f(X)= -\frac{1}{k!}\int_X^\infty du(X-u)^k \frac{d^{k+1}}{du^{k+1}} f(u)$$, 
$$\hspace{1.2cm}   = -\frac{X^{k+1}}{k!}\int_1^\infty dt(1-t)^k \partial^{k+1}_{(Xt)}f(Xt).$$ 
 Transcribed to arbitrary dimension and with $X = p$ this takes the form
\begin{equation}
 f(p) \equiv-(k+1){\displaystyle \sum_{\mid \beta \mid=k+1}}\big[\frac{p^{\beta}}{\beta !}\int^{\infty}_{1} \!\!dt (1-t)^{k} 
\partial^{\beta}_{(pt)}f(pt)\big]. 
\end{equation}
\vspace{0.5cm}
 It is easy to check that an equivalent representation of $f(p)$  is 
\vspace{-0.7cm}
\begin{equation}
f^{>}(p) \equiv-(k+1){\displaystyle \sum_{\mid \beta \mid=k+1}}\int^{\infty}_{1} \!\!dt\frac{(1-t)^{k}}{t^{(1-d)}} 
\partial^{\beta}_{(p)}\big[\frac{p^{\beta}}{\beta !}f^{>}(pt)\big]. 
\end{equation}
Here $d=D-1$. The same restriction as discussed earlier applies to  the $t$-integral. After partial integration in the UV-extension of $T(X)$ 
and in the limit $\alpha \rightarrow 1$ the upper limit in the $t$-integration is just $\mu^2$. The Fourier transform 
$\phi(x)$ to the conventionnal $x$-space representation  gives an alternative interpretation of Lagrange's formula, Eq.(IV.1), 
valid in the IR. Normalising the Fourier-integral with $(2 \pi)^{(D/2)}$ and after the change of integration variable  $t \rightarrow  1/t$ 
it reads
\vspace{-0.3cm}
\begin{equation}
\phi(x) =(k+1){\displaystyle \sum_{\mid \beta \mid=k+1}}\big[
\frac{x^{\beta}}{\beta !}\int^{1}_{0} dt \frac{(1-t)^{k}}{t^{(k+1)}}\partial^{\beta}_{(xt)}\phi(xt)\big]\},
\end{equation}
which is Eq.(IV.1). However when used in the analysis of a distribution singular in the IR and in the limit $\alpha \rightarrow 1$ 
the lower limit in the $t$-integration  comes from the  upper limit in the original $p$-integral and is now just $\frac{1}{\mu^2}.$
With this interpretation Eq.(IV.1) is particularly useful in establishing  generalized dispersion-type relations for causal distributions
in relation with the findings of \cite{Scharf}.


\begin{thebibliography}{10}
\bibitem{foote1}$f(p^2)$ is also $C^{\infty}$ with a fast decrease in the sense of L. Schwartz \cite{LSCW}.
\bibitem{foote2}for $d=1$ and $T(X)=\frac{1}{X}$  one has $\partial_X \log(\tilde{\mu} X )=Pf(\frac{1}{X})$ where $Pf$ is the 
pseudo-function distribution as defined in (\cite{LSCW}, chapt.II.2).
\end{thebibliography}
\end{document}